\documentclass[12pt]{article}
\usepackage{amssymb,amsmath,epsfig}
\begin{document} \parindent=0pt
\parskip=6pt \rm

\underline{\small To be publ., in: ``Progress in Ferromagnetic Research''
 (Nova Publ, N.Y., 2004).}

\vspace{0.5cm}
\begin{center}
 {\bf \Large Phases and phase
transitions in spin-triplet ferromagnetic superconductors}

{\bf  D. V. Shopova and D. I. Uzunov$^{\ast, \dag}$}

{\em  CPCM Laboratory, G. Nadjakov
Institute of Solid State Physics,\\
Bulgarian Academy of Sciences, BG-1784
Sofia, Bulgaria.} \\ \end{center}

$^{\ast}$ Also, Max-Planck-Institut f\"ur Physik komplexer Systeme,
N\"othnitzer Str. 38, 01187 Dresden, Germany.

$^{\dag}$  Corresponding author: uzun@issp.bas.bg

{\bf Key words}: superconductivity, ferromagnetism, phase diagram,\\
order parameter profile.

{\bf PACS}: 74.20.De, 74.20.Rp

\normalsize

\begin{abstract}
Recent results for the coexistence of ferromagnetism and
unconventional superconductivity with spin-tiplet Cooper pairing
are reviewed on the basis of the quasi-phenomenological
Ginzburg-Landau theory. New results are presented for the
properties of phases and phase transitions in such ferromagnetic
superconductors. The superconductivity, in particular, the mixed
phase of coexistence of ferromagnetism and unconventional
superconductivity is triggered by the spontaneous magnetization.
The mixed phase is stable whereas the other superconducting phases
that usually exist in unconventional superconductors are either
unstable, or, for particular values of the parameters of the
theory, some of these phases are metastable at relatively low
temperatures in a quite narrow domain of the phase diagram. The
phase transitions from the normal phase to the phase of
coexistence is of first order while the phase transition from the
ferromagnetic phase to the coexistence phase can be either of
first or second order depending on the concrete substance. The
Cooper pair and crystal anisotropies are relevant to a more
precise outline  of the phase diagram shape and  reduce the
degeneration of the ground states of the system but they do not
drastically influence the phase stability domains and the
thermodynamic properties of the respective phases. The results are
discussed in view of application to metallic ferromagnets as
UGe$_2$, ZrZn$_2$, URhGe.
\end{abstract}

{\bf \large 1. Introduction}

{\bf 1.1. Notes about unconventional superconductivity}

The phenomenon of unconventional Cooper pairing of fermions, i.e.
the formation of Cooper pairs with nonzero angular momentum was
theoretically predicted~\cite{Pitaevskii:1959} in 1959 as a
mechanism of superfluidity of Fermi liquids. In 1972 the same
phenomenon - unconventional superfluidity due to a $p$-wave (spin
triplet) Cooper pairing of $^3$He atoms, was experimentally
discovered in the mK range of temperatures; for details and
theoretical description, see
Refs.~\cite{Leggett:1975,Vollhardt:1990,Volovik:2003}. Note that,
in contrast to the standard $s$-wave pairing in usual
(conventional) superconductors, where the electron pairs are
formed by an attractive electron-electron interaction due to a
virtual phonon exchange, the widely accepted mechanism of the
Cooper pairing in superfluid $^3$He is based on an attractive
interaction between the fermions ($^3$He atoms) due to a virtual
exchange of spin fluctuations. Certain spin fluctuation mechanisms
of unconventional Cooper pairing of electrons have been assumed
also for the discovered in 1979 heavy fermion superconductors
(see, e.g., Refs.~\cite{Stewart:1984, Sigrist:1991, Mineev:1999})
as well as for some classes
 of high-temperature superconductors (see, e.g., Refs.~\cite{Sigrist:1987,
Annett:1988, Volovik:1988, Blagoeva:1990, Uzunov:1990, Uzunov:1993,
Annett:1995, Harlingen:1995, Tsuei:2000}).

The possible superconducting phases in unconventional
superconductors are described in the framework of the general
Ginzburg-Landau (GL) effective free energy
functional~\cite{Uzunov:1993} with the help of the  symmetry
groups theory. Thus a variety of possible superconducting
orderings were predicted for different crystal
structures~\cite{Volovik0:1985, Volovik:1985, Ueda:1985,
Blount:1985, Ozaki:1985, Ozaki:1986}. A detailed thermodynamic
analysis~\cite{Blagoeva:1990, Volovik:1985} of the homogeneous
(Meissner) phases and a renormalization group
investigation~\cite{Blagoeva:1990} of the superconducting phase
transition up to the two-loop approximation have been also
performed (for a three-loop renormalization group analysis, see
Ref.~\cite{Antonenko:1994}; for effects of magnetic fluctuations and disorder,
see~\cite{Busiello:1991, Busiello:1990}). We shall essentially use these
results in our present consideration.

In 2000, experiments~\cite{Saxena:2000} at low temperatures ($T
\sim 1$ K) and high pressure ($T\sim 1$ GPa) demonstrated the
existence of spin triplet superconducting states in the metallic
compound UGe$_2$. This superconductivity is triggered by the
spontaneous magnetization of the ferromagnetic phase which exists
at much higher temperatures and coexists with the superconducting
phase in the whole domain of existence of the latter below $T \sim
1$ K; see also experiments published in Refs.~\cite{Huxley:2001,
Tateiwa:2001}, and  the discussion in Ref.~\cite{Coleman:2000}.
Moreover, the same phenomenon of existence of superconductivity at
low temperatures and high pressure in the domain of the $(T,P)$
phase diagram where the ferromagnetic order is present has been
observed in other ferromagnetic metallic compounds
(ZrZn$_2$~\cite{Pfleiderer:2001} and  URhGe~\cite{Aoki:2001}) soon
after the discovery~\cite{Saxena:2000} of superconductivity in
UGe$_2$. 

In contrast to other superconducting materials, for
example, ternaty and Chevrel phase compounds, where the effects of
magnetic order on superconductivity are also substantial (see,
e.g.,~\cite{Vonsovsky:1982, Maple:1982, Sinha:1984, Kotani:1984}),
in these ferromagnetic compounds the phase transition temperature
($T_f$) to the ferromagnetic state is much higher than the phase
transition temperature ($T_{FS})$ from ferromagnetic to a (mixed) 
state of coexistence of ferromagnetism and superconductivity. For
example, in UGe$_2$ we have 
$T_{FS} = 0.8$ K whereas the critical temperature of the phase
transition from paramagnetic to ferromagnetic state in the 
same material is $T_f =35 $K~\cite{Saxena:2000, Huxley:2001}. One may 
reliably assume that in such kind of materials the material parameter
$T_s$ defined as the (usual) critical temperature
of the second order phase transition from normal to uniform (Meissner)
supercondicting state in zero external magnetic field is quite lower
than the phase transition temperature $T_{FS}$. Note, that the mentioned
 experiments on the compounds~UGe$_{2}$, URhGe, and ZrZn$_2$ 
do not give any evidence for the existence of
a standard normal-to-superconducting phase transition in zero external
magnetic field.

Moreover, it seems that the
superconductivity in the metallic compounds mentioned above,
always coexists with the ferromagnetic order and is enhanced by
the latter. As claimed in Ref.~\cite{Saxena:2000} in these systems
the superconductivity seems to arise from the same electrons that
create the band magnetism, and is most naturally understood as a
triplet rather than spin-singlet pairing phenomenon. Note that all
three metallic compounds, mentioned so far, are itinerant
ferromagnets. Besides, the unconventional superconductivity has
been suggested~\cite{Saxena:2001} as a possible outcome of recent
experiments in Fe~\cite{Shimizu:2001}, in which a superconducting
phase was discovered at temperatures below $2$ K at pressures
between 15 and 30 GPa. Note, that both vortex and Meissner
superconductivity phases~\cite{Shimizu:2001} are found in the
high-pressure crystal modification of Fe which has a hexagonal
close-packed lattice. In this hexagonal lattice the strong
ferromagnetism of the usual bcc iron crystal probably
disappears~\cite{Saxena:2001}. Thus one can hardly claim that
there is a coexistence of ferromagnetism and superconductivity in
Fe but the clear evidence for a superconductivity is also a
remarkable achievement.

{\bf 1.2. Ferromagnetism versus superconductivity}

The important point in all discussions of the interplay of
superconductivity and ferromagnetism is that a small amount of
magnetic impurities can destroy superconductivity in conventional
($s$-wave) superconductors by breaking up the ($s$-wave) electron
pairs with opposite spins (paramagnetic impurity
effect~\cite{Abrikosov:1960}). In this aspect the phenomenological
arguments~\cite{Ginzburg:1956} and the conclusions on the basis of
the microscopic theory of magnetic impurities in $s$-wave
superconductors~\cite{Abrikosov:1960} are in a complete agreement
with each other; see, e.g., Refs.~\cite{Vonsovsky:1982,
Maple:1982, Sinha:1984, Kotani:1984}. In fact, a total suppression
of conventional ($s$-wave) superconductivity should occur in the
presence of an uniform spontaneous magnetization
$\mbox{\boldmath$M$}$, i.e. in a standard ferromagnetic
phase~\cite{Ginzburg:1956}. The physical reason for this
suppression is the same as in the case of magnetic impurities,
namely, the opposite electron spins in the $s$-wave Cooper pair
turn over along the vector $\mbox{\boldmath$M$}$ in order to lower
their Zeeman energy and, hence, the pairs break down. Therefore,
the ferromagnetic order can hardly coexist with conventional
superconducting states. In particular, this is the case of
coexistence of uniform superconducting and ferromagnetic states
when the superconducting order parameter
$\psi(\mbox{\boldmath$x$})$ and the magnetization
$\mbox{\boldmath$M$}$ do not depend on the spatial vector
$\mbox{\boldmath$x$}$.

But yet a coexistence of $s$-wave superconductivity and
ferromagnetism may appear in uncommon materials and under quite
special circumstances. Furthermore, let us emphasize that
the conditions for the coexistence of
nonuniform (``vertex'', ``spiral'', ``spin-sinosoidal'' or ``helical'')
superconducting and ferromagnetic states are less restrictive than that for
the coexistence of uniform superconducting and ferromagnetic orders.
Coexistence of nonuniform phases has been discussed in details,
 in particular, experiment and theory of ternary and Chevrel-phase
compounds, where such a coexistence seems quite likely; for a
comprehensive review, see, for example, Refs.
~\cite{Vonsovsky:1982, Maple:1982, Sinha:1984,Kotani:1984,
Buzdin:1983}. 

In fact, the only two superconducting systems for
which the experimental data allow assumptions in a favor of a
coexistence of superconductivity and ferromagnetism are the rare
earth ternary boride compound ErRh$_4$B$_4$ and the Chervel phase
compound HoMo$_6$S$_8$; for a more extended review, see
Refs.~\cite{Maple:1982, Machida:1984}. In these compounds the
phase of coexistence most likely appears in a very narrow
temperature region just below the Curie temperature $T_f$ of the
ferromagnetic phase transition. At lower temperatures the magnetic
moments of the rare earth 4$f$ electrons become better aligned,
the magnetization increases and the $s$-wave superconductivity
pairs formed by the conduction electrons disintegrate.

{\bf 1.3. Unconventional superconductivity triggered by ferromagnetic order}

We shall not extend our consideration over all important aspects
of the long standing problem of coexistence of superconductivity
and ferromagnetism  rather we shall concentrate our attention on
the description of the newly discovered coexistence of
ferromagnetism and unconventional (spin-triplet) superconductivity
in the itinerant ferromagnets
 UGe$_2$, ZrZn$_2$, and URhGe. Here we wish to emphasize that the main object
of our discussion is the superconductivity of these compounds and,
at a second place in the rate of importance we put the problem of
coexistence. The reason is that the existence of superconductivity
in such itinerant ferromagnets is a highly nontrivial phenomenon.
As noted in Ref.~\cite {Machida:2001} the superconductivity in
these materials seems difficult to explain in terms of previous
theories~\cite {Vonsovsky:1982, Maple:1982, Kotani:1984} and seems
to require new concepts to interpret the experimental data.

We have already mentioned that in ternary
compounds the ferromagtetism comes from the localized 4$f$ electrons whereas
the s-wave Cooper pairs are formed by conduction electrons. In UGe$_2$ and
URhGe the 5$f$ electrons of U atoms form both superconductivity and
ferromagnetic order~\cite{Saxena:2000, Aoki:2001}. In ZrZn$_2$ the same
 double role is played by the 4$d$ electrons of Zr.
Therefore the task is to describe this behavior of the band
electrons at a microscopic level. One may speculate about a
spin-fluctuation mediated unconventional Cooper pairing as is in
case of $^3$He and heavy fermion superconductors. These important
issues have not yet a reliable answer and for this reason we shall
confine our consideration to a phenomenological level.

In fact, a number of reliable experimental data for example, the
data about the coherence length and the superconducting
gap~\cite{Saxena:2000, Huxley:2001, Aoki:2001, Pfleiderer:2001},
are in favor of the conclusion about a spin-triplet Cooper pairing
in these metallic compounds, although the mechanism of this
pairing remains unclear. We shall essentially use this reliable
conclusion. Besides, this point of view is consistent with the
experimental observation of coexistence of superconductivity only
in a low temperature part of the ferromagnetic domain of the phase
diagram ($T,P$), which means that a pure (non ferromagnetic)
superconducting phase has not been observed. This circumstance is
also in favor of the assumption of a spin-triplet
superconductivity.  Our investigation leads to results which confirm 
this general picture.

 Besides, on the basis of the experimental data
and conclusions presented for the first time in
Refs.~\cite{Saxena:2000, Coleman:2000} and shortly afterwards
confirmed in Refs.~\cite{Huxley:2001, Tateiwa:2001,
Pfleiderer:2001,
 Aoki:2001} one may reliably accept the point of view that the
the superconductivity in these magnetic compounds is
considerably enhanced by the ferromagnetic order parameter
$\mbox{\boldmath$M$}$ and, perhaps,
it could not exist without this ``mechanism of ferromagnetic trigger,'' or, in
short, ``$\mbox{\boldmath$M$}$-trigger.'' Such a
phenomenon is possible for spin-triplet Cooper pairs, where the electron
spins point parallel to each other and their turn along the vector of
the spontaneous magnetization $\mbox{\boldmath$M$}$ does not produce a
 break down of the spin-triplet Cooper pairs but rather stabilizes them and,
perhaps, stimulates their creation. We shall describe this phenomenon at a
phenomenological level.

{\bf 1.4. Phenomenological studies}

Recently, the  phenomenological theory which explains the
coexistence of ferromagnetism and unconventional spin-triplet
superconductivity of Landau-Ginzburg type was developed
~\cite{Machida:2001, Walker:2002}. The possible low-order
couplings between the superconducting and ferromagnetic order
parameters were derived with the help of general symmetry group
arguments and several important features of the superconducting
vortex state in the ferromagnetic phase of unconventional
ferromagnetic superconductors were established~\cite{Machida:2001,
Walker:2002}.

 In this article we shall use the approach presented
in Refs.~\cite{Machida:2001, Walker:2002}  to investigate
the conditions for the occurrence of the Meissner phase and
to demonstrate that the presence of ferromagnetic order enhances
the $p$-wave superconductivity. Besides, we shall establish
the phase diagram corresponding to model ferromagnetic
superconductors in a zero external magnetic field. We shall show
that the phase transition to the superconducting state in ferromagnetic
superconductors can be either of first or second order depending on
the particular substance. We confirm the predictions made in
Refs.~\cite{Machida:2001,Walker:2002} about the symmetry of the ordered
phases.

Our investigation is based on the mean-field
approximation~\cite{Uzunov:1993} as well as on familiar results
about the possible phases in nonmagnetic superconductors with triplet
($p$-wave) Cooper pairs~\cite{Volovik:1985, Blagoeva:1990,
Uzunov:1990}. Results from Refs.~\cite{Shopova1:2003, Shopova2:2003,
Shopova3:2003} will be reviewed and extended. In our preceding investigation
~\cite{Shopova1:2003, Shopova2:2003, Shopova3:2003} both
Cooper pair anisotropy and crystal
anisotropy have been neglected in order to clarify the main effect of the
coupling between the ferromagnetic and superconducting order parameters.
The phenomenological GL free energy is quite complex and the inclusion of
these anisotropies is related with lengthy formulae and a
multivariant analysis which
obscures the final results.

Here we shall take into account essential anisotropy effects, in
particular, the effect of the Cooper pair anisotropy on the
existence and stability of the mixed phase, namely the phase of
coexistence of superconductivity and ferromagnetic order. We
demonstrate that the anisotropy of the spin-triplet Cooper pairs
modifies but does not drastically change the thermodynamic
properties of this coexistence phase, in particular, in the most
relevant temperature domain above the superconducting critical
temperature $T_s$. The same is valid for the crystal anisotropy,
but we shall not present a thorough thermodynamic analysis of 
this problem. The crystal anisotropy effect can be considered
for concrete systems with various crystal
structures~\cite{Sigrist:1991, Volovik:1985}. Here we find enough
to demonstrate that the anisotropy is not crucial for the
description of the coexistence phase. Of course, our investigation
confirms the general concept~\cite{Volovik:1985} that the 
anisotropy reduces the degree of degeneration of the ground state
and, hence, stabilizes the ordering along the main crystal
directions.

There exists a formal similarity between the phase diagram
obtained in our investigation and the phase diagram of certain
improper ferroelectrics~\cite{Gufan:1980, Gufan:1981, Latush:1985,
Toledano:1987, Gufan:1987, Cowley:1980}. The variants of the
theory of improper ferroelectrics, known before 1980, were
criticized in Ref.~\cite{Cowley:1980} for their oversimplification
and inconsistency with the experimental results. But the further
development of the theory has no such disadvantage (see, e.g.,
Ref.~\cite{Toledano:1987, Gufan:1987}).
   We use the advantage of the theory of improper
ferroelectrics, where the concept of a ``primary'' order parameter
triggered by a secondary order parameter (the electric
polarization $\mbox{\boldmath$P$}_e$) has been initially
introduced and exploited (see Ref.~\cite{Toledano:1987,
Gufan:1987, Cowley:1980}). The mechanism of the M-triggered
superconductivity in itinerant ferromagnets is formally identical
to the mechanism of appearance of structural order triggered by
the electric polarization $\mbox{\boldmath$P$}_e$ in improper
ferroelectrics ($P$-trigger). Recently, the effect of $M$-trigger
has been used in a theoretical treatment of ferromagnetic Bose
condensates~\cite{Gu:2003}.

{\bf 1.5. Aims of the paper}

In the remainder of this paper we shall consider the GL free
energy functional of unconventional ferromagnetic superconductors.
Our aim is to establish the uniform phases which are described by
the GL free energy presented in Section 2.1. More information
about the justification of this investigation is presented in
Section 2.2. Note, as also mentioned in Section 2.2, that we
investigate a quite general GL model in a situation of a lack of a
concrete information about the values of the parameters of this
model for concrete compounds (UGe$_2$, URhGe, ZrZn$_2$) where the
ferromagnetic superconductivity has been discovered. On one side
this lack of information makes impossible a detailed comparison of
the theory to the available experimental data but on the other
side our results are not bound to one or more concrete substances
but can be applied to any unconventional ferromagnetic
superconductor. In Section 3 we discuss the phases in nonmagnetic
unconventional superconductors.
 In Section 4 the M-trigger effect will be described in the simple case of
 a single coupling (interaction)
between the magnetization $\mbox{\boldmath$M$}$
and the superconducting order parameter $\psi$ in an isotropic model of
ferromagnetic superconductors, where the anisotropy
effects are ignored. In Section 5 the effect of another important
coupling between the
magnetization and the superconducting order parameter on the thermodynamics of
the ferromagnetic superconductors is taken into
account. In Section 6 the anisotropy effects are considered. In Section 7 we
summarize and discuss our findings.

{\bf \large 2. Ginzburg-Landau free energy}

Following Refs.~\cite{Volovik:1985, Machida:2001, Walker:2002} in
this Chapter we discuss the phenomenological theory of
spin-triplet ferromagnetic superconductors and justify our
consideration in Sections 3--6.

{\bf 2.1. Model}

Consider the GL free energy functional
\begin{equation}
\label{eq1}
F[\psi,\mbox{\boldmath$M$}]=\int d^3 x f(\psi, \mbox{\boldmath$M$})\:,
\end{equation}
where the free energy density $f(\psi,\mbox{\boldmath$M$})$ (for
short hereafter called ``free energy'') of a spin-triplet
ferromagnetic superconductor is a sum of five terms:
\begin{equation}
\label{eq2}
f(\psi, \mbox{\boldmath$M$}) = f_{\mbox{\scriptsize S}}(\psi) +
f^{\prime}_{\mbox{\scriptsize
F}}(\mbox{\boldmath$M$}) +
f_{\mbox{\scriptsize I}}(\psi,\mbox{\boldmath$M$}) +
\frac{\mbox{\boldmath$B$}^2}{8\pi} - \mbox{\boldmath$B.M$}\:.
\end{equation}
In Eq.~(2) $\psi = \left\{\psi_j;j=1,2,3\right\}$ is the three-dimensional
complex vector describing the superconducting order and $\mbox{\boldmath$B$}
= (\mbox{\boldmath$H$} + 4\pi\mbox{\boldmath$M$}) = \nabla \times
\mbox{\boldmath$A$}$ is the magnetic induction; $\mbox{\boldmath$H$}$ is the
external magnetic field, $\mbox{\boldmath$A$} = \left\{A_j;
j=1,2,3\right\}$ is the magnetic vector potential. The last two terms on the
r.h.s. of Eq.~(2) are related with the magnetic energy which includes both
diamagnetic and paramagnetic effects in the
superconductor (see, e.g., \cite{Vonsovsky:1982, Ginzburg:1956, Blount:1979}).

In Eq.~(2), the term $f_{\mbox{\scriptsize S}}(\psi)$
describes the superconductivity for
$\mbox{\boldmath$H$} = \mbox{\boldmath$M$} \equiv 0$. This free energy part
can  be written in the form~\cite{Volovik:1985}
\begin{equation}
\label{eq3}
f_{\mbox{\scriptsize S}}(\psi)= f_{grad}(\psi)
 + a_s|\psi|^2 +\frac{b_s}{2}|\psi|^4 + \frac{u_s}{2}|\psi^2|^2 +
\frac{v_s}{2}\sum_{j=1}^{3}|\psi_j|^4 \;,
\end{equation}
with
\begin{eqnarray}
\label{eq4}
f_{grad}(\psi)&  =
& K_1(D_i\psi_j)^{\ast}(D_iD_j) +K_2\left[
 (D_i\psi_i)^{\ast}(D_j\psi_j) + (D_i\psi_j)^{\ast}(D_j\psi_i)\right] \\
\nonumber
&& + K_3(D_i\psi_i)^{\ast}(D_i\psi_i),
\end{eqnarray}
where a summation over the indices $i,j$ $(=1,2,3)$ is assumed and the symbol
\begin{equation}
\label{eq5}
 D_j = - i\hbar\frac{\partial}{\partial x_i} + \frac{2|e|}{c}A_j
\end{equation}
of covariant differentiation is introduced. In Eq.~(3), $b_s > 0$
and $a_s = \alpha_s(T-T_s)$, where $\alpha_s$ is a positive
material parameter and $T_s$ is the critical temperature of a
standard second order phase transition which may take place at $H
= {\cal{M}} = 0$; $H =|\mbox{\boldmath$H$}|$, and ${\cal{M}} =
|\mbox{\boldmath$M$}|$. The parameter $u_s$ describes the
anisotropy of the spin-triplet Cooper pair whereas the crystal
anisotropy is described by the parameter
$v_s$~\cite{Blagoeva:1990, Volovik:1985}. In Eq.~(3) the
parameters $K_j$, $(j = 1,2,3)$ are related with the effective
mass tensor of anisotropic Cooper pairs~\cite{Volovik:1985}.

The term
$f^{\prime}_{\mbox{\scriptsize F}}(\mbox{\boldmath$M$})$ in Eq.~(2)
is the following part of the free energy of a standard isotropic ferromagnet:
\begin{equation}
\label{eq6}
f^{\prime}_{\mbox{\scriptsize F}}(\mbox{\boldmath$M$}) =
c_f\sum_{j=1}^{3}|\nabla_j\mbox{\boldmath$M$}_j|^2 +
 a_f(T^{\prime}_f)\mbox{\boldmath$M$}^2 + \frac{b_f}{2}\mbox{\boldmath$M$}^4
\end{equation}
where $\nabla_j = \partial/\partial x_j$, $b_f > 0$, and
$a_f(T^{\prime}_f) = \alpha_f(T-T^{\prime}_f)$ is represented by
the material parameter $\alpha_f > 0$ and the temperature
$T^{\prime}_f$; the latter differs from the critical temperature
$T_f$ of the ferromagnet and this point will be discussed below.
In fact, through Eq.~(2) we have already added a negative term
($-2\pi {\cal{M}}^2$) to the total free energy
$f(\psi,\mbox{\boldmath$M$})$. This is obvious when we set $H = 0$
in Eq.~(2). Then we obtain the negative energy
($-2\pi{\cal{M}}^2$) which should be added to
$f^{\prime}_{\mbox{\scriptsize F}}(\mbox{\boldmath$M$})$. In this
way one obtains the total free energy $f_{\mbox{\scriptsize F}}
(\mbox{\boldmath$M$})$ of the ferromagnet in a zero external
magnetic field, which is given by a modification of Eq.~(6)
according to the rule
\begin{equation}
\label{eq7}
f_{\mbox{\scriptsize F}} (a_f) = f^{\prime}_{\mbox{\scriptsize F}}
\left[a_f(T^{\prime}_f) \rightarrow a_f(T_f) \right],
\end{equation}
where  $a_f = \alpha_f (T - T_f)$ and
\begin{equation}
\label{eq8}
T_f =  T^{\prime}_f + \frac{2\pi}{\alpha_f}
\end{equation}
is the critical temperature of a standard ferromagnetic phase
transition of second order. This scheme was used in studies of
rare earth ternary compounds~\cite{Vonsovsky:1982, Blount:1979,
Greenside:1981, Ng:1997}. Alternatively~\cite{Kuper:1980}, one may
work from the beginning with the total ferromagnetic free energy
$f_{\mbox{\scriptsize F}}(a_f,\mbox{\boldmath$M$})$ as given by
Eqs.~(6)~-~(8) but in this case the magnetic energy included in
the last two terms on the r.h.s. of Eq.~(2) should be replaced
with $H^2/8\pi$. Both ways of work are equivalent.

Finally, the term
\begin{equation}
\label{eq9}
f_{\mbox{\scriptsize I}}(\psi, \mbox{\boldmath$M$}) = i\gamma_0
\mbox{\boldmath$M$}.(\psi\times \psi^*) + \delta \mbox{\boldmath$M$}^2
|\psi|^2\;.
\end{equation}
in Eq.~(2) describes the interaction between the ferromagnetic
order parameter $M$ and the superconducting order parameter
$\psi$~\cite{Machida:2001,Walker:2002}. The  $\gamma_0$-term is
the most substantial for the description of experimentally found
ferromagnetic superconductors~\cite{Walker:2002} while the $\delta
\mbox{\boldmath$M$}^2 |\psi|^2$--term makes the model more
realistic in the strong coupling limit because it gives the
opportunity to enlarge the phase diagram including both positive 
and negative values of the parameter $a_s$. This allows for an extension of
the domain of the stable ferromagnetic order up to zero temperatures 
for a wide range of values of the material parameters and the pressure $P$.
Such a picture corresponds to the real situation in ferromagnetic compounds.

In Eq.~(9) the coupling constant
$\gamma_0 >0$ can be represented in the form $\gamma_0 = 4\pi J$,
where $J > 0$ is the ferromagnetic exchange parameter~\cite{Walker:2002}.
In general, the parameter $\delta$ for ferromagnetic superconductors may take
 both positive and negative values. The values of the material parameters
($T_s$, $T_f$, $\alpha_s$, $\alpha_f$,
$b_s$, $u_s$, $v_s$, $b_f$, $K_j$, $\gamma_0$ and $\delta$) depend on
the choice of the concrete substance and on intensive thermodynamic parameters,
such as the temperature $T$ and the pressure $P$.

{\bf 2.2. Way of treatment}

The total free energy (2) is a quite complex object of theoretical
investigation. The various vortex and uniform phases described by
this complex model cannot be investigated within a single
calculation but rather one should focus on concrete problems. In
Ref.~\cite{Walker:2002} the vortex phase was discussed with the
help of the criterion~\cite{Abrikosov:1957} for a stability of
this state near the phase transition line $T_{c2}(H)$; see also,
Ref.~\cite{Lifshitz:1980}. In case of $H = 0$ one should apply the
same criterion with respect to the magnetization ${\cal{M}}$ for
small values of $|\psi|$ near the phase transition line
$T_{c2}({\cal{M}})$ as performed in Ref.~\cite{Walker:2002}. Here
we shall be interested in the uniform phases, namely, when the
order parameters $\psi$ and $\mbox{\boldmath$M$}$ do not depend on
the spatial vector $\mbox{\boldmath$x$}\in V$ ($V$ is the volume
of the superconductor). Thus our analysis will be restricted to
the consideration of the coexistence of uniform (Meissner) phases
and ferromagnetic order. We shall perform this investigation in
details and, in particular, we shall show that the main properties
of the uniform phases can be given within an approximation in
which the crystal anisotropy is neglected. Moreover, some of the
main features of the uniform phases in unconventional
ferromagnetic superconductors can be reliably outlined when the
Cooper pair anisotropy is neglected, too.

The assumption of a uniform magnetization $\mbox{\boldmath$M$}$ is
always reliable outside a quite close vicinity of the magnetic
phase transition and under the condition that the superconducting
order parameter $\psi$ is also uniform, i.e. that vortex phases
are not present at the respective temperature domain. This
conditions are directly satisfied in type I superconductors but in
type II superconductors the temperature should be sufficiently low
and the external magnetic field should be zero. Moreover, the
mentioned conditions for type II superconductors may turn
insufficient for the appearance of uniform superconducting states
in materials with quite high values of the spontaneous
magnetization. In such cases the uniform (Meissner)
superconductivity and, hence, the coexistence of this
superconductivity with uniform ferromagnetic order may not appear
even at zero temperature. Up to now type I unconventional
ferromagnetic superconductors have not been yet found whereas the
experimental data for the recently discovered compounds UGe$_2$,
URhGe, and ZrZn$_2$ are not enough to conclude definitely either
about the lack or the existence of uniform superconducting states
at low and ultra-low temperatures. 

In all cases, if real materials can be described by the general GL free
energy (1)~-~(9), the ground state properties will be described by
uniform states, which we shall investigate. The problem about the
availability of such states in real materials at finite
temperatures is quite subtle at the present stage of research when
the experimental data are not enough. We shall assume that uniform
phases may exist in some unconventional ferromagnetic
superconductors. Moreover, we find convenient to emphasize that
these phases appear as solutions of the GL equations corresponding
to the free energy (1)~-~(9). These arguments completely justify
our study.

In case of a strong easy axis type of magnetic anisotropy, as is
in UGe$_2$~\cite{Saxena:2000}, the overall complexity of
mean-field analysis of the free energy $f(\psi,
\mbox{\boldmath$M$})$ can be avoided by performing an
``Ising-like'' description: $\mbox{\boldmath$M$} =
(0,0,{\cal{M}})$, where ${\cal{M}} = \pm |\mbox{\boldmath$M$}|$ is
the magnetization along the ``$z$-axis." Further, because of the
equivalence of the ``up'' and ``down'' physical states $(\pm
\mbox{\boldmath$M$})$ the thermodynamic analysis can be performed
within the ``gauge" ${\cal{M}} \geq 0$. But this stage of
consideration can also be achieved without the help of crystal
anisotropy arguments. When the magnetic order has a continuous
symmetry one may take advantage of the symmetry of  the total free
energy $f(\psi, \mbox{\boldmath$M$})$ and avoid the consideration
of equivalent thermodynamic states that occur as a result of the
respective symmetry breaking at the phase transition point but
have no effect on thermodynamics of the system. In the isotropic
system one may again choose a gauge, in which the magnetization
vector has the same direction as  $z$-axis ($|\mbox{\boldmath$M$}|
= M_z = {\cal{M}}$) and this will not influence the generality of
thermodynamic analysis. Here we shall prefer the alternative
description within which the ferromagnetic state may appear
through two equivalent ``up'' and ``down'' domains with
magnetizations $ {\cal{M}}$ and ($ -{\cal{M}}$), respectively.

We shall perform the mean-field analysis of the uniform phases and
the possible phase transitions between such phases in a zero
external magnetic field ($\mbox{\boldmath$H$}=0)$, when the
crystal anisotropy is neglected ($v_s \equiv 0$). The only
exception will be the consideration
 in Sec.~3, where we briefly discuss the nonmagnetic superconductors
(${\cal{M}} \neq 0$). For our aims we use notations in which the number of
 parameters is reduced. Introducing the parameter
\begin{equation}
\label{eq10}
b = (b_s + u_s + v_s)
\end{equation}
we redefine the order parameters and the other parameters in the following way:
\begin{eqnarray}
\label{eq11}
&&\varphi_j =b^{1/4}\psi_j = \phi_je^{\theta_j}\:,\;\;\;
M = b_f^{1/4}{\cal{M}}\:,\\ \nonumber
&& r = \frac{a_s}{\sqrt{b}}\:,\;\;\; t =\frac{a_f}{\sqrt{b_f}}\:, \;\;\;
w = \frac{u_s}{b}\:, \;\;\; v =\frac{v_s}{b}\:, \\ \nonumber
&&\gamma= \frac{\gamma_0}{b^{1/2}b_f^{1/4}}\:,\;\;\;
\gamma_1= \frac{\delta}{(bb_f)^{1/2}}\:.
\end{eqnarray}

Having in mind our approximation of uniform $\psi$ and $\mbox{\boldmath$M$}$
 and the notations~(10)~-~(11), the free energy density
 $f(\psi,M) = F(\psi,M)/V$ can be written in the form
\begin{eqnarray}
\label{eq12}
f(\psi,M)& = & r\phi^2 + \frac{1}{2}\phi^4
  + 2\gamma\phi_1\phi_2 M \mbox{sin}(\theta_2-\theta_1) + \gamma_1 \phi^2 M^2
+ tM^2 + \frac{1}{2}M^4\\ \nonumber
&& -2w \left[\phi_1^2\phi_2^2\mbox{sin}^2(\theta_2-\theta_1)
 +\phi_1^2\phi_3^2\mbox{sin}^2(\theta_1-\theta_3) +
 \phi_2^2\phi_3^2\mbox{sin}^2(\theta_2-\theta_3)\right] \\ \nonumber
&& -v[\phi_1^2\phi_2^2 + \phi_1^2\phi_3^2 + \phi_2^2\phi_3^2].
\end{eqnarray}
Note, that in this free energy the order parameters $\psi$ and
$\mbox{\boldmath$M$}$ are defined per unit volume.

The equilibrium phases are obtained from the equations of state
\begin{equation}
\label{eq13}
\frac{\partial f(\mu_0)}{\partial \mu_{\alpha}} = 0\:,
 \end{equation}
 where the series of symbols $\mu$ can be defined as, for example,
 $\mu = \left\{\mu_\alpha\right\}=
 (M, \phi_1,..., \phi_3,$ $ \theta_1,..., \theta_3)$; $\mu_0$ denotes an
equilibrium phase.
The stability matrix $\tilde{F}$ of the phases $\mu_0$
 is defined by
\begin{equation}
\label{eq14}
 \hat{F}(\mu_0)= \left\{F_{\alpha\beta}(\mu_0)\right\} = \frac{\partial^2f(\mu_0)}
{\partial\mu_{\alpha}\partial\mu_{\beta}}\;.
\end{equation}

An alternative treatment can be done in terms of the real
($\psi^{\prime}_j$) and imaginary ($\psi^{\prime\prime}_j$) parts
of the complex numbers $\psi_j = \psi_j^{\prime} +
i\psi_j^{\prime\prime}$. The calculation with the moduli $\phi_j$
and phase angles $\theta_j$ of $\psi_j$  has a minor disadvantage
in cases of strongly degenerate phases when some of the angles
$\theta_j$ remain unspecified. Then one should consistently use
the properties of the respective broken continuous symmetry.
Alternatively, one may do an alternative analysis with the help of
the components $\psi_j^{\prime}$ and $\psi_j^{\prime\prime}$.

In order to avoid any ambiguity in our discussion let us note that
we often use the term ``existence'' of a phase in order to
indicate that it appears in experiments. This means that the
phase, we consider, is either stable or metastable, in quite rare
cases, when certain special special experimental conditions allow
 the observation of metastable states in equilibrium. When a
solution (phase) of Eq.~(13) is obtained it is  said that the
respective phase ``exists'', of course, under some ``existence
conditions'' that are imposed on the parameters $\left\{
\mu_{\alpha} \right\}$ of the theory. But this is just a
registration of the fact that a concrete phase satisfies Eq.~(13). 

The problem about the thermodynamic stability of the phases that
are solutions of Eq.~(13) is solved with the help of the matrix
(14) and, if necessary, with an additional analysis including the
comparison of the free energies of phases which correspond to
minima of the free energy in one and the same domain of parameters
$\{\mu_{\alpha}\}$. Then the stable phase will be the phase that
corresponds to a global minimum of the free energy. Therefore,
when we discuss experimental situation in which some phase exists
according to the experimental data, this means that it is a global
minimum of the free energy, a fact determined by a comparison of
free energies of the phases. If other minima of the free energy
exist in a certain domain of parameters $\{\mu_{\alpha}\}$ then
these minima are metastable equilibria, i.e. metastable phases. If
a solution of Eq.~(13) is not a minimum, it corresponds to an
(absolutely) unstable equilibrium and the matrix (14)
corresponding to this unstable phase is negatively definite. 

When we determine the minima of the free energy by the requirement for
a positive definiteness of the stability matrix (14), we are often
faced with the problem of a ``marginal'' stability, i.e. the
matrix is neither positively nor negatively definite. This is
often a result of the degeneration of the states (phases) with
broken continuous symmetry, and one should distinguish these
cases. If the reason for the lack of a clear positive definiteness
of the stability matrix is precisely the mentioned degeneration of
the ground state, one may reliably conclude that the respective
phase is stable. If there is another reason, the analysis of the
matrix (14) turns insufficient for our aims to determine the
respective stability property. These cases are quite rare and
happen for very particular values of the parameters
$\{\mu_{\alpha}\}$.

{\bf 3. Pure superconductivity}

Let us set $M\equiv 0$ in Eq.~(12) and  briefly summarize the known
 results~\cite{Volovik:1985,Blagoeva:1990} for the
 ``pure superconducting case'' when the magnetic order cannot appear and
 magnetic effects do not affect the
 stability of the normal and uniform (Meissner) superconducting phases. The 
possible phases can be classified by the structure of the
complex vector order parameter $\psi = (\psi_1,\psi_2,\psi_2)$. We
shall often use the moduli vector $(\phi_1, \phi_2,\phi_3)$ with
magnitude $\phi = (\phi_1^2+\phi_2^2+\phi_3^2)^{1/2}$ but we must
not forget the values of the phase angles $\theta_j$.

The normal phase (0,0,0) is always a solution of the Eqs.~(13). It
is stable for $r\geq 0$, and corresponds to a free energy $f=0$.
Under certain conditions, six ordered
phases~\cite{Volovik:1985,Blagoeva:1990} occur for $r<0$. Here we
shall not repeat the detailed description of these
phases~\cite{Volovik:1985, Blagoeva:1990} but we shall briefly
mention their structure.

The simplest ordered phase is of type $(\psi_1,0,0)$ with
equivalent domains: $(0,\psi_2,0)$ and $(0,0,\psi_3)$. Multi-
domain phases of more complex structure also occur, but we shall
not always enumerate the possible domains. For example, the
``two-dimensional'' phases
 can be fully represented by domains of type
 $(\psi_1,\psi_2,0)$ but there are also other two types of domains:
 $(\psi_1,0,\psi_3)$ and
$(0,\psi_2,\psi_3)$. As we consider the general case when the crystal
anisotropy is present $(v \neq 0)$, this type of phases possesses the property
 $|\psi_i| = |\psi_j|$.

The two-dimensional phases are two and have different free
energies. To clarify this point let us consider, for example, the
phase $(\psi_1,\psi_2,0)$. The two complex numbers, $\psi_1$ and
$\psi_2$ can be represented either as two-component real vectors,
or, equivalently, as rotating vectors in the complex plane. One
can easily show that Eq.~(12) yields two phases: a collinear
phase, when $(\theta_2-\theta_1) = \pi k (k = 0,\pm1,...)$, i.e.
when the vectors $\psi_1$ and $\psi_2$ are collinear, and another
(noncollinear) phase when the same vectors are perpendicular to
each other: $(\theta_2-\theta_1) = \pi(k + 1/2)$. Having in mind
that $|\phi_1| = |\phi_2| = \phi/\sqrt{2}$, the domain
$(\psi_1,\psi_2,0)$ of the collinear phase is given by
$(\pm1,1,0)\phi/\sqrt{2}$, and the same domain for the
 noncollinear phase is given by $(\pm i,1,0)\phi/\sqrt{2}$.
Similar representations can be given for the other two domains
of these phases.

In addition to the mentioned three ordered phases, three other
ordered phases exist. For these phases all three components
$\psi_j$ have nonzero equilibrium values.  Two of them have equal
to one another moduli $\phi_j$, i.e., $\phi_1=\phi_2=\phi_3$. The
third phase is of the type $\phi_1=\phi_2 \neq \phi_3$  and is
unstable so it cannot occur in real systems. The two
three-dimensional phases with equal moduli of the order parameter
components have different phase
 angles and, hence, different structure. The difference between any couple
 of angles $\theta_j$ is given by $\pm \pi/3$ or $\pm 2\pi/3$. The
characteristic vectors of this phase can be of the form
$(e^{i\pi/3}, e^{-i\pi/3},1)\phi/\sqrt{3}$ and $(e^{2i\pi/3},
 e^{-i2\pi/3},1)\phi/\sqrt{3}$. The second stable three dimensional phase
 is ``real'', i.e. the components $\psi_j$ lie on the real axis;
 $(\theta_j-\theta_j) = \pi k$ for any couple of angles $\theta_j$ and the
characteristic vectors are $(\pm 1, \pm 1, 1)\phi/\sqrt{3}$. The
stability properties of these five stable ordered phases were
presented in details in Refs.~\cite{Volovik:1985,Blagoeva:1990}.

When the crystal anisotropy is not present ($v =0$) the picture
changes. The increase of the level of degeneracy of the ordered
states leads to an instability of some phases and to a lack of
some noncollinear phases. Both two- and three-dimensional real
phases, where $(\theta_j -\theta_j) = \pi k$, are no more
constrained by the condition $\phi_i=\phi_j$ but rather have the
freedom of a variation of the moduli $\phi_j$ under the condition
$\phi^2 = -r >0$. The two-dimensional noncollinear phase exists
but has a marginal stability~\cite{Blagoeva:1990}. All other
noncollinear phases even in the presence of a crystal anisotropy
$(v\neq 0)$ either vanish or are unstable; for details, see
Ref.~\cite{Blagoeva:1990}. This discussion demonstrates that the
crystal anisotropy stabilizes the ordering along the main
crystallographic directions, lowers the level of degeneracy of the
ordered state related with the spontaneous breaking of the
continuous symmetry and favors the appearance of noncollinear
phases.

The crystal field effects related to the unconventional
superconducting order were established for the first time in
Ref.~\cite{Volovik:1985}. In our consideration of unconventional
ferromagnetic superconductors in Sec.~4--7 we shall take advantage
of these effects of the crystal anisotropy. In both cases $v=0$
and $v \neq 0$ the matrix (14) indicates an
 instability of three-dimensional phases (all $\phi_j \neq 0)$ with an
 arbitrary ratios $\phi_i/\phi_j$. As already mentioned,
for $v \neq 0$ the phases of type $\phi_1= \phi_2\neq \phi_3$
 are also unstable whereas for $v=0$, even the phase $\phi_1=\phi_2=\phi_3 > 0$
is unstable.

{\bf 4. Simple case of M-triggered superconductivity}

Here we consider the Walker-Samokhin model~\cite{Walker:2002} when
 only the $M\phi_1\phi_2-$coupling between the
order parameters $\psi$ and $M$ is taken into account ($\gamma > 0$,
 $\gamma_1 = 0$). Besides, we shall neglect the anisotropies $(w=v=0)$.
The uniform phases and the phase diagram in this case were
investigated in Refs.~\cite{Shopova1:2003,Shopova2:2003,
Shopova3:2003}. Here we summarize the main results in order to
make a clear comparison with the new results presented in
Sections~5 and 6. In this Section we set $\theta_3 \equiv 0$ and
use the notation $\theta \equiv \Delta\theta = (\theta_2 -
\theta_1)$. The symmetry of the system allows to introduce the
notations without a loss of generality of the consideration.

{\bf 4.1. Phases}

The possible (stable, metastable and unstable)
 phases are given in Table 1 together with the respective
existence and stability conditions. The stability conditions define the
 domain of the phase diagram where the respective phase is either stable
or metastable~\cite{Uzunov:1993}. The normal (disordered) phase,
denoted in Table 1 by $N$ always exists (for all temperatures $T
\geq 0)$ but is stable for $t >0$, $r > 0$. The superconductivity
phase denoted in Table 1 by SC1 is unstable. The same is valid for
the phase of coexistence of ferromagnetism and superconductivity
denoted in Table 1 by CO2. The N--phase, the ferromagnetic phase
(FM), the superconducting phases (SC1--3) and two of the phases of
coexistence (CO1--3) are generic phases because they appear also in
the decoupled case $(\gamma\equiv 0)$. When the
$M\phi_1\phi_2$--coupling is not present, the phases SC1--3 are
identical and represented by the order parameter $\phi$ where the
components $\phi_j$ participate on  equal footing. The asterisk
attached to the stability condition of ``the second
superconductivity phase"(SC2), indicates that our analysis is
insufficient to determine whether this phase corresponds to a
minimum of the free energy. As we shall see later the phase SC2,
as well as the other two purely superconducting phases and the
coexistence phase CO1, have no chance to become stable for $\gamma
\neq 0$. This is so, because the non-generic phase of coexistence
of superconductivity and ferromagnetism (FS in Table 1), which
does not exist for $\gamma = 0$ is stable and has a lower free
energy in their domain of stability. Note, that a second domain
$(M < 0)$ of the FS phase exists and is denoted in Table 1 by
FS$^*$. Here we shall describe only the first domain (FS). The
domain FS$^{\ast}$ is considered in the same way.

The cubic equation for $M$ corresponding to FS (see Table~1) is
shown in Fig.~1 for $\gamma = 1.2$ and $t = -0.2$. For any $\gamma
> 0$ and $t$, the stable FS thermodynamic states are given by $r
(M) < r_m = r(M_m)$ for $M > M_m > 0$, where $M_m$ corresponds to
the maximum of the function $r(M)$. Functions $M_m(t)$ and $M_0(t)
= (-t + \gamma^2/2)^{1/2} = \sqrt{3}M_m(t)$ are drawn in Fig.~2
for $\gamma = 1.2$.  Functions $r_m(t) = 4M_m^3(t)/\gamma$ for $t
< \gamma^2/2$ (the line of circles in Fig.~3) and $r_e(t) =
\gamma|t|^{1/2}$ for $t < 0$  define the borderlines of stability
and existence of FS.

\vspace{0.3cm}

\small
TABLE 1. Phases and their existence and stability properties
 [$\theta = (\theta_2-\theta_1)$,
$k = 0, \pm 1,...$].\\
\begin{tabular}{|l|l|l|l|} \hline \hline
\small Phase & order parameter & existence & stability domain \\
\hline N & $\phi_j = M = 0$ & always & $t > 0, r > 0$ \\ \hline
 FM & $\phi_j = 0$, $M^2 = -t$& $t < 0$& $r>0$, $r^2 > \gamma^2t$\\ \hline
SC1 & $\phi_1=M=0$, $\phi^2 = -r$ & $r<0$ & unstable  \\ \hline SC2 &
$\phi^2 = -r$, $\theta = \pi k$, $M = 0$ & $r<0$ & $(t > 0)^*$
 \\ \hline
SC3 & $\phi_1=\phi_2=M=0$, $\phi^2_3 = -r$ & $r<0$
&$r<0$, $t>0$\\ \hline
CO1 &$\phi_1= \phi_2=0$, $\phi^2_3 = -r$, $M^2=-t$&$r<0$, $t<0$ &
$r<0$, $t<0$ \\ \hline
CO2 &$\phi_1=0$, $\phi^2 = -r$, $\theta=\theta_2=\pi k$, $M^2=-t$&
$r<0$, $t<0$ &  unstable \\ \hline
FS & $2\phi_1^2 = 2\phi_2^2 = \phi^2 = -r + \gamma M$, $\phi_3 = 0$ &
$\gamma M > r$ &  $3M^2>(-t +\gamma^2/2)$ \\
& $\theta= 2\pi(k - 1/4) $, 
$\gamma r = (\gamma^2-2t)M - 2M^3$ & & $M > 0$ \\
 \hline
FS$^{\ast}$ & $2\phi_1^2 = 2\phi_2^2 = \phi^2 = -(r + \gamma M)$, 
$\phi_3 = 0$ &
$-\gamma M > r$ &  $3M^2>(-t +\gamma^2/2)$ \\
& $\theta= 2\pi(k + 1/4) $, 
$\gamma r = (2t -\gamma^2)M + 2M^3$ & & $M < 0$ \\
 \hline \hline
\end{tabular}

\vspace{0.3cm}

\begin{figure}
\begin{center}
\epsfig{file=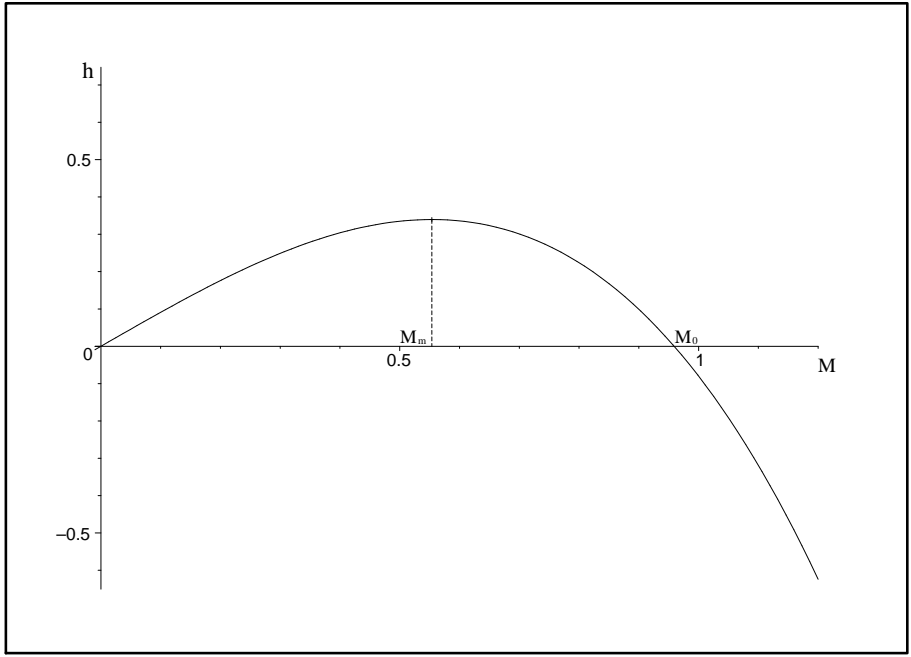,angle=-90, width=12cm}
\end{center}
\caption{$h=\gamma r/2$ as a function of $M$ for $\gamma = 1.2$, and $t
= -0.2$.} \label{Uzunovf1.fig}
\end{figure}
\begin{figure}
\begin{center}
\epsfig{file=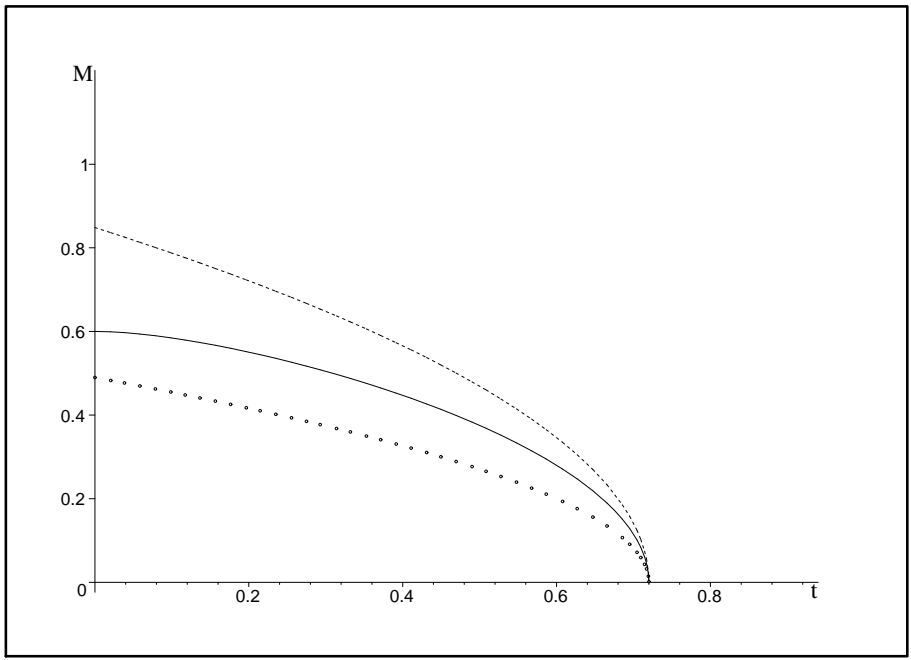,angle=-90, width=12cm}
\end{center}
\caption{$M$ versus $t$ for $\gamma = 1.2$: the dashed line represents
$M_0$, the solid line represents $M_{eq}$, and the dotted line
corresponds to $M_m$.} \label{Uzunovf2.fig}
\end{figure}

\normalsize

{\bf 4.2. Phase diagram}

We have outlined the domain in the ($t$, $r$) plane where
the FS phase exists and is a
 minimum of the free
energy. For $r < 0$ the cubic equation  for $M$ (see Table 1) and
the existence and stability conditions are satisfied for any $M
\geq 0$ provided $t \geq \gamma^2 $. For $ t < \gamma^2$ the
condition $M \geq M_0$ have to be fulfilled, here the value
 $M_0 = (-t + \gamma^2/2)^{1/2}$ of $M$ is obtained from $r(M_0) = 0$. Thus
for $r = 0$ the N-phase is stable for
 $t \geq \gamma^2/2$, on the other hand FS is stable for $t \leq \gamma^2/2$.
For $r > 0$, the requirement for the stability of FS leads to the
inequalities
\begin{equation}
\label{eq15}
  max\left(\frac{r}{\gamma}, M_m\right) < M < M_0\;,
\end{equation}
where $M_m = (M_0/\sqrt{3})$ and $M_0$ should be the positive
solution of the cubic equation of state from Table~1; $M_m > 0$
gives a maximum of the function $r(M)$; see also Figs.~1 and 2.
\begin{figure}
\begin{center}
\epsfig{file=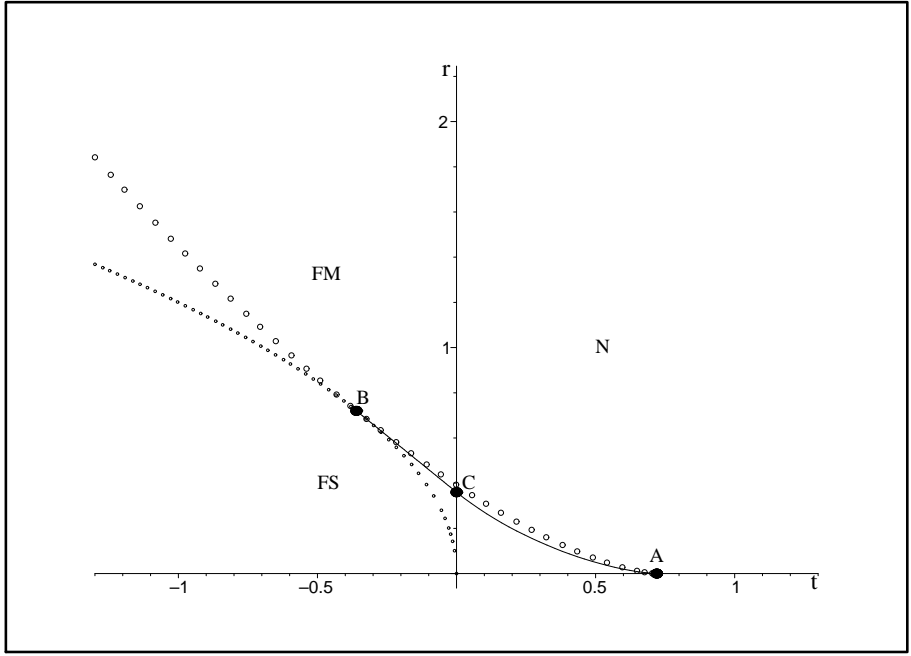,angle=-90, width=12cm}\\
\end{center}
\caption{The phase diagram in the plane ($t$, $r$) with two tricritical
points (A and B) and a triple point $C$; $\gamma = 1.2$. The domains of
existence and stability of the phases N, FM and FS are shown.}
 \label{Uzunovf3.fig}
 \end{figure}

The further analysis leads to the existence and  stability domain of FS
below the line AB given by circles (see Fig.~3). In Fig.~3 the curve of
circles starts from the point A with coordinates ($\gamma^2/2$, $0$)
and touches two other (solid and dotted) curves at the point B with
coordinates ($-\gamma^2/4$, $\gamma^2/2$).  Line of
 circles represents the function
$r(M_m) \equiv r_m(t)$ where
\begin{equation}
\label{eq16}
 r_m(t) = \frac{4}{3\sqrt{3}\gamma} \left (\frac{\gamma^2}{2} -
 t\right)^{3/2}.
\end{equation}
Dotted line is given by $r_e(t) = \gamma\sqrt{|t|}$. The inequality $r
< r_m(t)$ is a condition for the stability of FS, whereas the
inequality $r \leq r_e(t)$ for $ (-t) \geq \gamma^2/4$ is a condition
for the existence of FS as a solution of the respective equation of
state. This existence condition for FS has been obtained from $\gamma M
> r$ (see Table 1).

In the region on the left of the point B in Fig.~3, the FS phase
satisfies the existence condition $\gamma M > r$ only
 below the dotted line. In the domain confined between the lines of circles
 and the dotted
line on the left of the point B the stability condition for FS is
satisfied but the existence condition is broken. The inequality $r \geq
r_e(t)$ is the stability condition of FM for $ 0 \leq (-t) \leq
\gamma^2/4$. For $(-t) > \gamma^2/4$ the FM phase is stable for all $r
\geq r_e(t)$.

In the region confined by the line of circles AB, the dotted line
for $ 0 < (-t) < \gamma^2/4$, and the $t-$axis, the phases N, FS
and FM have an overlap of stability domains. The same is valid for
FS, the SC phases and CO1 in the third quadrant of the plane ($t$,
$r$). The comparison of the respective free energies for $r < 0$
shows that the stable phase is FS whereas the other phases are
metastable within their domains of stability.

The part of
 the $t$-axis given by $r=0$ and $t > \gamma^2/2$
 is a phase transition line of second order
which describes the N-FS transition. The same transition
 for $0 < t < \gamma^2/2$ is represented by the solid line AC which
is the equilibrium transition line of a first order phase transition.
This equilibrium transition curve is given by the function
\begin{equation}
\label{eq17}
 r_{eq}(t) =
\frac{1}{4}\left[3\gamma - \left(\gamma^2 + 16t
\right))^{1/2}\right]M_{eq}(t),
\end{equation}
where
  \begin{equation}
\label{eq18}
 M_{eq}(t) =
\frac{1}{2\sqrt{2}}\left[\gamma^2 - 8t + \gamma\left(\gamma^2 +
 16t \right)^{1/2}\right]^{1/2}
\end{equation}
is the equilibrium value (jump) of the magnetization. The order of
the N-FS transition changes at the tricritical point A.

The domain above the solid line AC and below the line of circles for $
t > 0$ is the region of a possible
 overheating of FS.
The domain of overcooling of the N-phase is confined by the solid line AC and
the axes ($t > 0$, $r >0$). At the triple point C with coordinates
 [0, $r_{eq}(0) = \gamma^2/4$]
the phases N, FM, and FS
coexist. For $t < 0$ the straight line
\begin{equation}
\label{eq19}
r_{eq}^* (t) =  \frac{\gamma^2}{4} + |t|,\;\;\;\;\;\; -\gamma^2/4 < t < 0,
\end{equation}
describes the extension of the equilibrium phase transition line of the
N-FS first order transition to negative values of $t$.
 For $t < (-\gamma^2/4)$
 the equilibrium phase transition FM-FS is of second order and is
given by the dotted line on the left of the point B (the second
tricritical point in this phase diagram). Along the first order
transition line
 $r_{eq}^{\ast}(t)$ given by~ Eq.~(\ref{eq19}) the equilibrium value
 of $M$ is $M_{eq} =\gamma/2$,  which
implies an equilibrium order parameter jump at the FM-FS transition equal to
($\gamma/2 - \sqrt{|t|}$). On the dotted line of the second order
FM-FS
 transition the equilibrium value
of $M$ is equal to that of the FM phase ($M_{eq} = \sqrt{|t|}$).
Note, that the FM phase does not exists below $T_s$ and this seems
to be a disadvantage of the model~(\ref{eq12}) with $\gamma_1 =
0$.

The equilibrium phase transition lines of the FM-FS and N-FS transition
lines in Fig.~3 can be expressed by the respective equilibrium phase
transition temperatures $T_{eq}$ defined by the equations $r_e =
r(T_{eq})$, $r_{eq} = r(T_{eq})$, $r^{\ast}_{eq} = r(T_{eq})$, and with
the help of the relation $M_{eq} = M(T_{eq})$. This leads to some
limitations on the possible variations of the parameters of the theory.
 For example, the critical temperature
($T_{eq} \equiv T_c$) of the FM-FS transition of second order
 ($\gamma^2/4 < -t$)  is obtained in the form
$T_{c} = (T_s + 4\pi J{\cal{M}}/\alpha_s)$,
or, using ${\cal{M}} = (-a_f/\beta)^{1/2}$,
\begin{equation}
\label{eq20}
T_{c} = T_s -\frac{T^{\ast}}{2} +
\left[ \left(\frac{T^{\ast}}{2}\right)^2 + T^{\ast}(T_f-T_s)\right]^{1/2}\;,
\end{equation}
where $T_f > T_s$, and $T^{\ast} = (4\pi J)^2\alpha_f/\alpha_s^2\beta$ is
 a characteristic temperature of the model~(\ref{eq12}) with $\gamma_1=w=v=0$.
The investigation of the conditions for the validity of Eq.~(\ref{eq20})
leads to the conclusion that the FM-FS continuous phase transition
(at $\gamma^2 < -t)$ is possible only if the following condition is satisfied:
\begin{equation}
\label{eq21}
T_{f} - T_s > \ = (\varsigma + \sqrt{\varsigma})T^{\ast}\;,
\end{equation}
where $\varsigma = \beta\alpha_s^2/4b\alpha_f^2$.
 This means that the second
order FM-FS transition should disappear for a sufficiently large
$\gamma$--coupling. Such a condition does not exist for the first order
transitions FM-FS and N-FS.

Taking into  account the gradient term (4) in the free energy~(\ref{eq2})
should lead to a depression of the equilibrium transition temperature.
As the magnetization increases with the decrease of the temperature,
the vortex state should occur at temperatures which are lower than
 the equilibrium temperature $T_{eq}$ of
the homogeneous (Meissner) state. For example,
the critical temperature ($\tilde{T}_c$)
 corresponding to the inhomogeneous (vortex) phase
of FS-type has been evaluated~\cite{Walker:2002} to be
 lower than the critical temperature ~(\ref{eq20}): $(T_c - \tilde{T}_c) =
4\pi \mu_B{\cal{M}}/\alpha_s$ ($\mu_B = |e|\hbar/2mc$ - Bohr magneton).
 For $J \gg \mu_B$, we have $T_c \approx \tilde{T}_c$.

For $ r > 0$, namely, for temperatures $T > T_s$ the
superconductivity is triggered by the magnetic order through the
$\gamma$-coupling. The superconducting phase for $T > T_s$ is
entirely in the $(t,r)$ domain of the ferromagnetic phase.
Therefore, the uniform supeconducting phase can occur for $T >
T_s$ only through a coexistence with the ferromagnetic order.

In the next Sections we shall focus
on the temperature range $T > T_s$ which seems to be of main practical
interest. We shall not dwell on the superconductivity in the fourth quadrant
 $(t >0,r<0)$ of the $(t,r)$ diagram where pure superconductivity phases
are possible in systems with $T_s > T_f$ (this is not the case for UGe$_2$,
URhGe, and ZrZn$_2$). Besides, we shall not discuss the
possible metastable phases in the third quadrant $(t<0,r<0)$ of the
$(t,r)$ diagram.

{\bf 4.3. Magnetic susceptibility}

Consider the longitudinal magnetic susceptibility $\chi_1 =
(\chi_{\mbox{\scriptsize V}}/V)$ per unit
volume~\cite{Shopova3:2003}. The external magnetic field
$\mbox{\boldmath$H$} = (0,0,H)$ with $ H = \left(\partial
f/\partial {\cal{M}}\right)$ has the same direction as the
magnetization $\mbox{\boldmath$M$}$. We shall calculate the
quantity $\chi = \sqrt{\beta_f}\chi_1$ for the equilibrium
thermodynamic states $\mu_0$ given by Eq.~(13). Having in mind the
relations (11) between $M$ and ${\cal{M}}$, and between $\psi$ and
$\varphi$ we can write
\begin{equation}
\label{eq22} \chi^{-1} =  \frac{d}{d M_0}\left[\left(\frac{\partial f}
{\partial M}\right)_{T,\varphi_j} \right]_{\mu_0}\:,
\end{equation}
where the equilibrium magnetization $M_0$ and equilibrium
superconducting order parameter components $\varphi_{0j}$ should
be taken for the respective equilibrium phase (see Table~1, where
the suffix ``0'' of $\phi$, $\theta$, and $M$ has been omitted;
hereafter the same suffix will be often omitted, too). Note that
the value of the equilibrium magnetization $M$ in FS is the
maximal nonnegative root of the cubic equation in $M$ given in
Table~1.

Using Eq.~(22) we obtain the susceptibility $\chi$ of
the FS phase in the form
\begin{equation}
\label{eq23} \chi^{-1} = -\gamma^2 +  2t + 6M^2\;.
\end{equation}
The susceptibility of the other phases has the usual expression
\begin{equation}
\label{eq24} \chi^{-1} =  2t + 6M^2\;.
\end{equation}
Eq.~(24) yields the known results for the paramagnetic
susceptibility
  ($\chi_P = 1/2t$; $t>0$) , corresponding to the normal phase, and for the
ferromagnetic susceptibility ($\chi_F = 1/4|t|$; $t <0$), corresponding to FM.
These susceptibilities can be compared with the susceptibility $\chi$ of FS.
As the susceptibility $\chi$ of FS cannot be analytically calculated for the
whole domain of stability of FS, we shall consider the close vicinity of
the N-FS and FM-FS phase transition lines.

Near the second order phase transition line on the left of the point
$B$ ($t < -\gamma^2/4$), the magnetization has a smooth behaviour and
the magnetic susceptibility does not exhibit any singularities (jump or
divergence). For $t > \gamma^2/2$, the magnetization is given by $M =
(s_- + s_+)$, where
\begin{equation}
\label{eq25}
 s_{\pm} =\left\{- \frac{\gamma r}{4} \pm \left[
\frac{(t-\gamma^2/2)^3}{27} + \left( \frac{\gamma
r}{4}\right)^2\right]^{1/2} \right\}^{1/3}\;.
\end{equation}
For $r = 0$, $M = 0$, whereas for $|\gamma r| \ll (t - \gamma^2/2)$ and
$r=0$ one may obtain $M \approx -\gamma r/ (2t-\gamma^2) \ll 2t$. This
means that in a close vicinity $(r < 0)$ of $r = 0$ along the second
order phase transition line $(r=0, t>\gamma^2)$ the magnetic
susceptibility is well described by the paramagnetic law $\chi_P =
(1/2t)$. For $r< 0$ and $t \rightarrow \gamma^2/2$, we obtain $M =
-(\gamma r/2)^{1/3}$ which yields
\begin{equation}
\label{eq26}
 \chi^{-1} =  6\left(\frac{\gamma |r|}{2}\right)^{2/3}\:.
\end{equation}

On the phase transition line $AC$ we have
\begin{equation}
\label{eq27} M_{eq}(t) = \frac{1}{2\sqrt{2}}\left[\gamma^2 - 8t +
\gamma\left(\gamma^2 + 16t \right)^{1/2}\right]^{1/2}
\end{equation}
and, hence,
\begin{equation}
\label{eq28}
 \chi^{-1} = -4t - \frac{\gamma^2}{4}\left[1 -3 \left( 1 +
 \frac{16t}{\gamma^2}\right)^{1/2}\right]\:.
\end{equation}
At the tricritical point $A$ this result yields $\chi^{-1}(A) = 0$,
whereas at the triple point $C$ with coordinates ($0$, $\gamma^2/4$) we
have $\chi(C) = (2/\gamma^2)$. On the line $BC$ we obtain
$M=\gamma/2$ and, hence,
\begin{equation}
\label{eq29} \chi^{-1} = 2t + \frac{\gamma^2}{2}\:.
\end{equation}
At the tricritical point $B$ with coordinates ($-\gamma^2/4$,
$\gamma^2/2$) this result yields $\chi^{-1}(B)= 0$.

In order to investigate the magnetic susceptibility tensor
we shall slightly extend the framework of out treatment by considering
arbitrary orientations of the vectors $\mbox{\boldmath$H$}$ and
$\mbox{\boldmath$M$}$. We shall denote the spatial directions
$(\mbox{\boldmath$x$},\mbox{\boldmath$y$},\mbox{\boldmath$z$})$ as $(1,2,3)$.

The components of the inverse magnetic susceptibility tensor
\begin{equation}
\label{eq30}
\hat{\chi}^{-1}_1 =\hat{\chi}^{-1}\sqrt{b_f} = \left\{\chi^{-1}_{ij}\right\}
\sqrt{b_f}
\end{equation}
can be represented in the form
\begin{equation}
\label{eq31}
\chi^{-1}_{ij} = 2(t + M^2)\delta_{ij} + 4M_iM_j +
i\gamma\frac{\partial}{\partial M_j}(\varphi\times\varphi^{\ast})_i\:,
\end{equation}
where $M$ and $\varphi_j$ are to be taken at their equilibrium values:
$M_0$, $\varphi_{0j}$, $\theta_{0j}$. The last term in the r.h.s. of Eq.~(28)
is equal to zero for all phases in Table~1 except for FS (and FS$^{\ast})$.
When the last term in Eq.~(29) is equal to zero we obtain the known
result the susceptibility tensor for second order phase transitions
(see, e.g., Ref.~\cite{Uzunov:1993}).

Consider the FS phase, where $\phi_{j}$
depends on $M_j$. Now we can choose again
$\mbox{\boldmath$M$} = (0,0,M)$ and use our results for the equilibrium values
of $\phi_j$, $\theta$ and $M$ (see Table~1). Then the
components $\chi^{-1}_{ij}$ corresponding to FS are given by
\begin{equation}
\label{eq32}
\chi^{-1}_{ij} = 2(t + M^2)\delta_{ij} + 4M_iM_j -\gamma^2\delta_{i3}\:.
\end{equation}
Thus we have $\chi^{-1}_{i\neq j}= 0$,
\begin{equation}
\label{eq33}
\chi^{-1}_{11} = \chi^{-1}_{22} = 2(t + M^2)\:,
\end{equation}
and $\chi^{-1}_{33}$  coincides with the inverse longitudinal susceptibility
 $\chi^{-1}$ given by  Eq.~(23).

 {\bf 4.4. Entropy and specific heat}

The entropy $S(T) \equiv (\tilde{S}/V)
= -V\partial (f/\partial T) $ and the specific heat $C(T) \equiv
(\tilde{C}/V) = T(\partial S/\partial T)$ per unit volume $V$ are
calculated in a standard way~\cite{Uzunov:1993}. We are interested in the
jumps of
these quantities on the N-FM, FM-FS, and N-FS transition lines. The
behaviour of $S(T)$ and $C(T)$ near the N-FM phase transition and near
the FM-FS phase transition line of second order on the left of the
point $B$ (Fig.~3) is known from the standard theory of critical
phenomena (see, e.g., Ref.~\cite{Uzunov:1993} and for this reason we focus our
attention on the phase transitions of type FS-FM and FS-N for
$(t>-\gamma^2/4$), i.e., on the right of the point $B$ in Fig.~3.

Using the equations for the order parameters $\psi$ and $M$ (Table~1) and
applying the standard procedure for the calculation of $S$, we obtain
the general expression
\begin{equation}
\label{eq34}
 S(T) = - \frac{\alpha_s}{\sqrt{b}}\phi^2 -
\frac{\alpha_f}{\sqrt{\beta}}M^2\:.
\end{equation}
The next step is to calculate the entropies $S_{ FS}(T)$ and
$S_{FM}$ of the ordered phases FS and FM. Note, that use the usual
convention $F_{N} = Vf_{N}=0$ for the free energy of the N-phase
and, hence, we must set $S_{N}(T)=0$.

Consider the second order phase transition line ($r=0$,
$t>\gamma^2/2$). Near this line $S_{FS}(T)$ is a smooth function of $T$
and has no jump but the specific heat $C_{FS}$ has a jump at $T=T_s$,
i.e. for $r=0$. This jump is given by
\begin{equation}
\label{eq35}
 \Delta C_{FS}(T_s) = \frac{\alpha_s^2T_s}{b}\left[ 1 -
 \frac{1}{1 - 2t(T_s)/\gamma^2}\right]\:.
\end{equation}
The jump $\Delta C_{FS}(T_s)$ is higher than the usual jump $\Delta
C(T_c) = T_c\alpha^2/b$ known from the Landau theory of standard second
order phase transitions~\cite{Uzunov:1993}.

The entropy jump $\Delta S_{AC}(T) \equiv S_{FS}(T) $ on the line $AC$
is obtained in the form
\begin{equation}
\label{eq36}
 \Delta S_{AC}(T) = -M_{eq}\left\{\frac{\alpha_s\gamma}{4\sqrt{b}}\left[1
 + \left(1 + \frac{16t}{\gamma^2}\right)^{1/2}\right] -
 \frac{\alpha_f}{\sqrt{\beta}}M_{eq}\right\}\:,
\end{equation}
where $M_{eq}$ is given by Eq.~(18). From Eqs.~(18) and (36), we have $\Delta
S(t=\gamma^2/2) = 0$, i.e., $\Delta S(T)$ becomes equal to zero at the
tricritical point $A$. Besides we find from Eqs.~(18) and (36) that at the
triple point $C$ the entropy jump is given by
\begin{equation}
\label{eq37} \Delta S(t=0) =
  -\frac{\gamma^2}{4}\left(\frac{\alpha_s}{\sqrt{b}} +
   \frac{\alpha_f}{\sqrt{\beta}} \right)\:.
\end{equation}

On the line $BC$ the entropy jump is defined by $\Delta S_{BC}(T) =
[S_{FS}(T)-S_{FM}(T)]$. We obtain
\begin{equation}
\label{eq38} \Delta S_{BC}(T) =
 \left( |t| -\frac{\gamma^2}{4}\right)\left(\frac{\alpha_s}{\sqrt{b}}
  + \frac{\alpha_f}{\sqrt{\beta}}
   \right)\:.
\end{equation}
At the tricritical point $B$ this jump is equal to zero as it should
be. The calculation of the specific heat jump on the first order phase
transition lines $AC$ and $BC$ is redundant for two reasons. Firstly,
the jump of the specific heat at a first order phase transition differs
from the entropy by a factor of order of unity. Secondly, in caloric
experiments where the relevant quantity is the latent heat $Q = T
\Delta S(T)$, the specific heat jump can hardly be distinguished.

{\bf 4.5. Note about a simplified theory}

The consideration in this Section as well as in Sections 5 and 6
can be performed within an approximate scheme, known from the
theory of improper ferroelectrics (see, e.g., Ref.~\cite{Cowley:1980}).
The idea of the approximation is in the supposition that the order
parameter $M$ is small enough so that one can neglect $M^4$-term
in the free energy. Within this approximation one easily obtains
from the data for FS presented in Table~1 or by a direct
calculation of the respective reduced free energy that the order
parameters $\phi$ and $M$ of FS are described by the simple
equalities $r = (\gamma M -\phi^2)$ and $M = (\gamma/2t)\phi^2$.
Of course, one may perform this simple analysis from the very beginning. 
For ferroelectrics this approximation gives a substantial
departure of theory from experiment~\cite{Cowley:1980}. In
general, the domain of reliability of such an approximation should
be the close vicinity of the ferromagnetic phase transition, i.e.
temperatures near to the critical temperature $T_f$. On the other
hand, this discussion is worthwhile only if the ``primary'' order
parameter also exists in the same (narrow) temperature domain
($\phi > 0$). Therefore this approximation has some application in
systems, where $T_s \ge T_f$.

 For $T_s<T_f$, one may 
simplify our thorough analysis by a supposition for a relatively small 
value of the modulus $\phi$ of the superconducting order parameter. 
This approximation should be valid in some narrow temperature domain near 
the line of second order phase transition from FM to FS.

{\bf 5. Effect of symmetry conserving coupling}

Here we consider the case when both coupling parameters $\gamma$
and $\gamma_1$ are different from zero. In this way we shall
investigate the effect of the symmetry conserving $\gamma_1$-term
in the free energy on the thermodynamics of the system. Note that
when $\gamma$ is equal to zero the analysis is quite easy and the
results are known from the theory of bicritical and tetracritical
points~\cite{Uzunov:1993, Toledano:1987, Liu:1973, Imry:1975}. For
the problem of coexistence of conventional superconductivity and
ferromagnetic order this analysis $(\gamma = 0, \gamma_1 \neq 0)$
was made in Ref.~\cite{Vonsovsky:1982}. Once again we postpone the
consideration of anisotropy effects by setting $w = v = 0$. The
present analysis is much more difficult than that in Sec.~4, and
cannot be performed only by analytical calculations; rather, some
complementary numerical analysis is needed. Our investigation is
based to a great extent on analytical calculations but a numerical
analysis has been also performed in order to obtain concrete
conclusions.

{\bf 5.1. Phases}

The calculations show that for temperatures $T > T_s$, i.e., for $r > 0$,
we have three stable phases. Two of them are quite simple:
the normal ($N$-) phase with existence and
stability domains shown in Table~1, and the
FM phase with the existence condition $ t<0$ as shown in Table~1, and
a stability domain defined by the inequalities $r > \gamma_1t$ and
\begin{equation}
\label{eq39}
r > \gamma_1t + \gamma\sqrt{-t}\:.
\end{equation}
The third stable phase for $r>0$ is a more complex variant of the mixed
phase FS and its domain FS$^*$, discussed in Section 4.
The symmetry of the FS phase coincides with that found in ~\cite{Walker:2002}

Let us also mention that for $r<0$ five pure superconducting ($M
=0$, $\phi > 0$) phases exist. Two of these phases, $(\phi_1 > 0,
\phi_2 = \phi_3 =0)$ and $(\phi_1 =0, \phi_2>0, \phi_3>0)$ are
unstable. Two other phases, $(\phi_1>0, \phi_2>0, \phi_3 =0,
\theta_2 = \theta_1 + \pi k)$ and $(\phi_1>0,\phi_2>0, \phi_3>0,
\theta_2 = \theta_1 + \pi k, \theta_3$ -- arbitrary; $k=0,\pm1,...)$
show a marginal stability for $ t > \gamma_1 r$. 

Only one of the
five pure superconducting phases, namely, the phase SC3, given in
Table~1, is stable. In the present case of $\gamma_1 \neq 0$ the
values of $\phi_j$ and the existence domain of SC3 are the same as
shown in Table ~1 for $\gamma_1 =0$ but the stability domain is
different and is given by $t > \gamma_1 r$. When the anisotropy
effects are taken into account the phases exhibiting marginal
stability within the present treatment may receive a further
stabilization. Besides, three other mixed phases $(M \neq, \phi
>0)$ exist for $r < 0$ but one of them is metastable (for
$\gamma_1^2 >1, t < \gamma_1 r$, and $r < \gamma_1 t$) and the
other two are absolutely unstable. Here the thermodynamic
behaviour for $r < 0$ is much more abundant in phases than in the
case of improper ferroelectrics with two component primary order
parameter ~\cite{Toledano:1987}. However, at this stage of
experimental needs about the properties of unconventional
ferromagnetic superconductors the investigation of the phases for
temperatures $T < T_s$ is not of primary interest and for this
reason we shall focus on the relatively higher temperature domain
$r > 0$.

The FS phase is described by the following equations:
\begin{equation}
\label{eq40}
\phi_1 = \phi_2=\frac{\phi}{\sqrt{2}}\:, \;\;\; \phi_3 = 0\:,
\end{equation}
\begin{equation}
\label{eq41}
\phi^2= (\pm \gamma M-r-\gamma_1 M^2)\:,
\end{equation}
\begin{equation}
\label{eq42}
(1-\gamma_1^2)M^3\pm \frac{3}{2} \gamma \gamma_1 M^2
+\left(t-\frac{\gamma^2}{2}-\gamma_1 r\right)M \pm \frac{\gamma
r}{2}=0\:,
\end{equation}
and
\begin{equation}
\label{eq43}
(\theta_2 - \theta_1) = \mp \frac{\pi}{2} + 2\pi k\:,
\end{equation}
($k = 0, \pm 1,...$). The upper sign in Eqs.~(41) - (43) corresponds to
the FS domain in which $\mbox{sin}(\theta_2-\theta_1) = -1$ and the lower sign
corresponds to the FS$^{*}$ domain. Here we have a generalization of
the two-domain phase FS discussed in Section 4 and for this reason we use
the same notations. The analysis of the stability matrix (14) for these phase
 domains shows that FS is stable for $M > 0$ and FS$^{*}$ is stable for
$M<0$, just like our result in Section 4. As these domains belong to the same
phase, namely, have the same free energy and are thermodynamically equivalent,
we shall consider one of them, for example, FS. Besides, our analysis of
Eqs.~(40) - (43) shows that FS exists and is stable in a broad
 domain of the $(t,r)$ diagram, including substantial regions corresponding to
$r>0$.

{\bf 5.2. Phase stability and phase diagram}

In order to outline the phase diagram ($t,r$) we shall use the
 information given above for the other three phases which have their own
 domains of stability in the $(t,r)$ plane: N, FM, and FS.
 \begin{figure}
 \begin{center}
 \epsfig{file=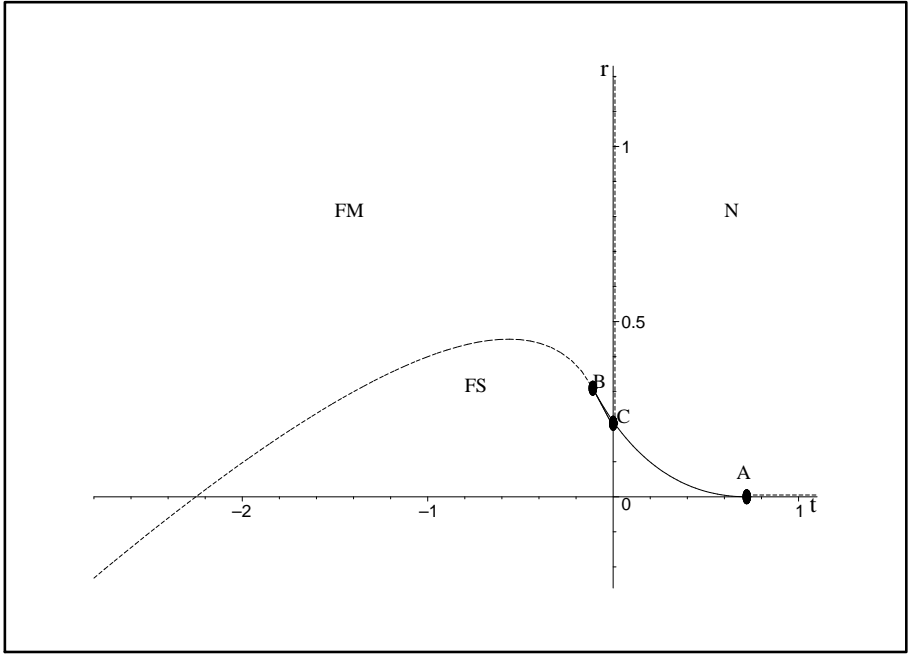,angle=-90, width=12cm}\\
 \end{center}
 \caption{The phase diagram in the $(t,r)$ plane for
 $\gamma=1.2,\;\gamma_1=0.8$ and $w=0$.} \label{Uzunovf4.fig}
 \end{figure}
The phase diagram for concrete parameters of $\gamma$ and $\gamma_1$ is shown
in Fig.~4. The phase transition between the normal and FS phases is of first
order and goes along the equilibrium line AC. It is given by the
equation:

\begin{equation}
\label{eq44}
r_{eq}(t)= \frac{M_{eq}}{(\gamma_1
M_{eq}-\gamma/2)}\left[(1-\gamma_1^2)M_{eq}^2+
\frac{3}{2} \gamma \gamma_1 M_{eq} +(t-\frac{\gamma^2}{2})\right].
\end{equation}
The equilibrium value $M_{eq}$ on the line AC is found by setting
the equilibrium free energy $f_{FS}(\mu_0)$ of FS equal to zero,
i.e. equal to the free energy ($f_N = 0$) of the N-phase. We have
obtained the equilibrium energy $f_N$ as a function of the
magnetization:
\begin{eqnarray}
\label{eq45}
f_{FS} & = & -\frac{M^2}{2(M\gamma_1-\gamma/2)^2}\\ \nonumber
&&\times\left\{(1-\gamma_1^2)M^4 + \gamma\gamma_1 M^3 +
2\left[t(1-\gamma_1^2)-
\frac{\gamma^2}{8}\right] M^2 - 2\gamma\gamma_1t M + 
t(t-\frac{\gamma^2}{2})\right\},
 \end{eqnarray}
where $M \equiv M_{eq}$ (hereafter the suffix ``eq'' will be often
omitted). 

The numerical analysis of the free energy (45) as a
polynomial of $M$ shows that the expression in the curly brackets
has one positive zero in the interval of values of $t$ from
$t=\gamma^2/2$ (point A in Fig.~4) up to $t=0$, where
$M_{t=0}=\gamma/2(\gamma_1+1)$. As far as the obtained values for
$M$ are in the interval $0 \le$M$<(\gamma/2 \gamma_1)$ the
existence condition of FS, namely,
\begin{equation}
\label{eq46}
\phi^2=\frac{M(M^2+t)}{(\gamma/2-\gamma_1 M)} \ge 0\:,
\end{equation}
is also satisfied.

At the triple point C with coordinates $t=0$, $r=\gamma^2/4(\gamma_1+1)$ three
phases (N, FM, and FS) coexist. To find the magnetization $M$ on the
equilibrium curve BC of the first order phase transition FM-FS for
t$<0$, we use the equality f$_{FM}$=f$_{FS}$, or, equivalently,
\begin{equation}
\label{eq47}
  \frac{(M^2+t^2)^2}{2(M\gamma_1-\gamma/2)^2}\left[\frac{\gamma}{2}-
M(1+\gamma_1)\right]
  \left[\frac{\gamma}{2}+M(1-\gamma_1)\right]=0.
\end{equation}
Then the function $r_{eq}$(t) for $t<0$ will have the form
\begin{equation}
\label{eq48}
  r_{eq}(t) = \frac{\gamma^2}{4(1+\gamma_1)}-t,
\end{equation}
This function describes the line BC of first order phase transition
(see Fig.~4) which terminates at the tricritical point B with coordinates

\begin{equation}
\label{eq49}
t_B = -\frac{\gamma^2}{4(1+\gamma_1)^2}\:,\;\;\; r_B=
\frac{\gamma^2(2+\gamma_1)}{4(1+\gamma_1)^2}\:.
\end{equation}
 To the left of the tricritical point the second order phase
transition curve is given by the relation,
\begin{equation}
\label{eq50}
r_e(t)=\gamma\sqrt{-t}+\gamma_1 t,
\end{equation}
which coincides with the stability condition (39) of FM. This
line intersects t-axis for $t=(-\gamma^2/\gamma_1^2)$ and is well
defined also for $r<0$. On the curve $r_e(t)$ the magnetization is
$M=\sqrt{-t}$ and the superconducting order parameter is equal to zero
 ($\phi=0$). The function $r_e(t)$ has
a maximum at the point $(t,r) = (-\gamma^2/4\gamma_1^2, \gamma^2/4\gamma_1)$;
here $M=(\gamma/2\gamma_1)$. When this point is approached the second
derivative of
the free energy with respect to $M$ tends to infinity,
but as we shall see later the inclusion of the anisotropy of triplet pairing
smears this singularity. The result for the curves $r_{FS}(t)$ of equilibrium
phase transitions (N-FS ans FM-FS) can be used to define the respective
equilibrium phase transition temperatures $T_{FS}$.

We shall not discuss the region, $t>0$, $r<0$, because we have
supposed from the very beginning of our analysis that the
transition temperature for the ferromagnetic order T$_f$ is higher
then the superconducting transition temperature T$_s$, as i is for
the known unconventional ferromagnetic superconductors. But this
case may become of substantial interest when, as one may expect,
materials with $T_f < T_s$ will be discovered.

 \begin{figure}
 \begin{center}
 \epsfig{file=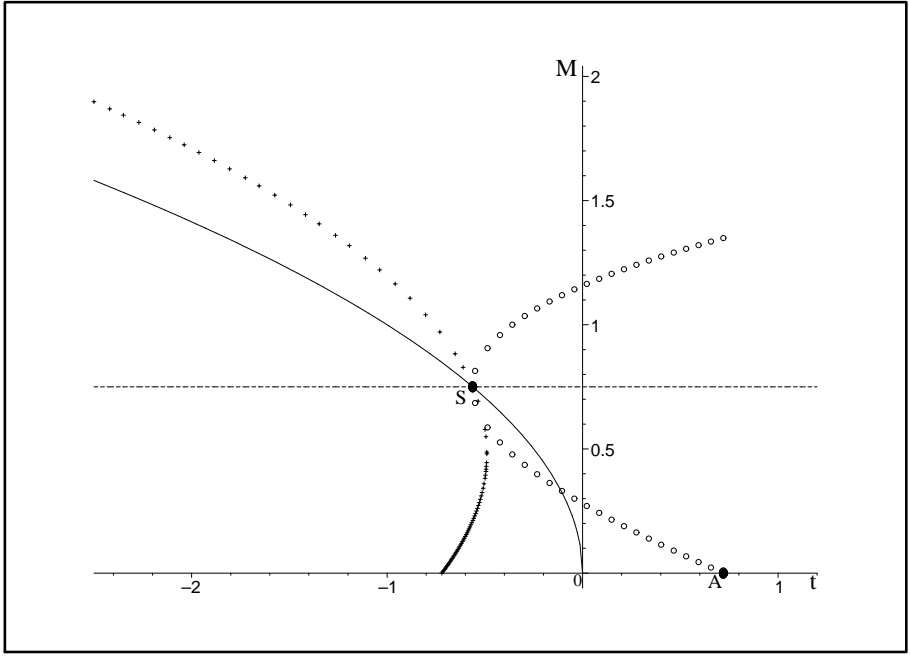,angle=-90, width=12cm}\\
 \end{center}
 \caption{The dependence $M(t)$ as an illustration of stability analysis for
 $\gamma=1.2$ and $\gamma_1=0.8$.} \label{Uzunovf5.fig}
 \end{figure}

The stability conditions of FS can be written in the general form
\begin{equation}
\label{eq51}
\frac{-M^2+\gamma \gamma_1 M
-t-\gamma^2/2}{M\gamma_1-\gamma/2}\ge0\:,
\end{equation}
\begin{equation}
\label{eq52}
\gamma M \ge 0\:,
\end{equation}
\begin{equation}
\label{eq53}
\frac{1}{M\gamma_1-\gamma/2}\left[\gamma_1(1-\gamma_1^2)M^3 -
\frac{3}{4}\gamma(1-2\gamma_1^2) M^2 -
\frac{3}{4}\gamma^2\gamma_1 M - \frac{\gamma}{4}(t-\gamma^2/2)\right]
\ge 0\:.
\end{equation}

Our consideration of the stability conditions~(51)~-~(53)~together
with the existence condition Eq.~(46) of the phase FS is
illustrated by the picture shown in Fig.~5. 

For $0 \le t \le
\gamma^2/2$ and $0<M<(\gamma/2\gamma_1)$ conditions (46) and
(51) are  satisfied. Condition (53) is a cubic equation in $M(t)$
which for the above values of the parameter $t$ has three real
roots, one of them negative. The positive roots, $M(t) > 0$, as
function of $t$ are drawn by circles in Fig. 5 and it is obvious
that the condition (53) will be satisfied for those values of
$M(t)$ that are between the two circled curves. The smaller
positive root of Eq.~(53) intersects $t$-axis for $t=\gamma^2/2$
(point A in Fig.~5). Note, that $M=\gamma/(2\gamma_1)$ is given by
the horizontal dashed line. For $t \le -\gamma^2(2-\gamma_1^2)/4$
the stability condition (51) has two real roots shown by curves
with crosses in Fig.~5. For negative values of the parameter $t$
we shall consider also the curve $M=\sqrt{-t}$ which is the
solution of existence condition (46) and is depicted by solid line
in Fig.~5. For $(-\gamma^2/4 \gamma_1^2)<t<0$ the FS phase exists
and is stable when $\gamma /(2 \gamma_1) \ge M\ge  \sqrt{-t}$.

The point S in Fig.~5 with coordinates $(-\gamma^2/(4 \gamma_1^2),
\gamma/(2 \gamma_1)$ is singular in sense that l.h.s. of
conditions (51) and (53) go to infinity there. When
$t>(-\gamma^2/4 \gamma_1^2)$ the existence condition (46) implies
$\gamma/(2 \gamma_1)<M<\sqrt{-t}$. The stability condition (53) is
always satisfied (two complex conjugate roots and one negative
root) and condition (51) will be fulfilled for values of $M$
between the two curves denoted by crosses in Fig.~5.

{\bf 5.3. Discussion}

The shape of the equilibrium phase transition lines corresponding
to the phase transitions N-SC, N-FS, and FM-FS is similar to that
for the simpler case $\gamma_1 = 0$ and we shall not dwell on the
variation of the size of the phase domains with the variations of
the parameter $\gamma_1$ from zero to values constrained by the
condition $\gamma_1^2 <1$. Besides one may generalize our
treatment (Section 4) of the magnetic susceptibility tensor and
the thermal quantities in this more complex case and to
demonstrate the dependence of these quantities on $\gamma_1$. We
shall not dwell on these problems. But an important qualitative
difference between the equilibrium phase transition lines shown in
Figs.~1 and 4 cannot be omitted. The second order phase transition
line $r_e(t)$, shown by the dotted line on the left of point $B$
in Fig.~1, tends to large positive values of $r$ for large
negative values of $t$ and remains in the ``second quadrant''
($t<0, r>0)$ of the plane ($t,r$) while the respective second
order phase transition line in Fig.~4 crosses the $t$-axis in the
point $t=-\gamma^2/\gamma_1^2$ and is located in the third
quadrant ($t<0,r<0$) for all possible values $t <
-\gamma^2/\gamma_1^2$. This means that the ground state (at 0 K)
of systems with $\gamma_1 =0$ will be always the FS phase whereas
two types of ground states, FM and FS,
 are allowed for systems with $0< \gamma_1^2 < 1$. The latter
seems more realistic in view of comparison of theory and
experiment, especially, in ferromagnetic compounds like UGe$_2$,
URhGe, and ZrZn$_2$. The neglecting of the $\gamma_1$-term does
not allow to describe the experimentally observed presence of FM
phase at quite low temperatures and relatively low pressure $P$.

The final aim of the phase diagram investigation is the outline of
the ($T,P$) diagram. Important conclusions about the shape of the
$(T,P)$ diagram can be made from the form of the $(t,r)$ diagram
without an additional information about the values of the relevant
material parameters $(a_s$, $a_f,...$) and their dependence on the
pressure $P$. One should know also the characteristic temperature
$T_s$, which has a lower value than the experimentally
observed~\cite{Saxena:2000, Huxley:2001, Tateiwa:2001,
Pfleiderer:2001, Aoki:2001} phase transition temperature $(T_{FS}
\sim 1 K)$ to the mixed (FS) phase. A supposition about the
dependence of the parameters $a_s$ and $a_f$ on the pressure $P$
was made in Ref.~\cite{Walker:2002}. Our results for $T_f \gg T_s$ show 
that the phase transition temperature $T_{FS}$ varies with the 
variation of the system parameters $(\alpha_s, \alpha_f,...)$ from values
which are much higher than the charactestic temperature $T_s$ up to
zero temperature. This is seen from Fig.~4.

{\bf 6. Anisotropy effects}

When the anisotropy of the Cooper pairs is taken in consideration,
there will be not drastic changes in the shape the phase diagram
for $r>0$ and the order of the respective phase transitions.
 Of course, there will be some changes in the size of the phase domains and
the formulae for the thermodynamic quantities. The parameter $w$
will also insert a slight change in the values of the
thermodynamic quantities like the magnetic susceptibility and the
entropy and specific heat jumps at the phase transition points.

Besides, and this seems to be the main anisotropy effect, the $w$-
and $v$-terms in the free energy lead to a stabilization of the
order along the main crystal directions which, in other words,
means that the degeneration of the possible ground states (FM, SC,
and FS) is considerably reduced. This means also a smaller number
of marginally stable states which are encountered by the analysis
of the definiteness of the stability matrix (14). All anisotropy
effects can be verified by the investigation of the free energy
(12) which includes the $w$- and $v$-terms. 

We have made the above
general conclusions on the basis of a detailed analysis of the
effect of the Cooper pair anisotropy ($w$-) term, as well as on
the basis of a preliminary analysis of the total free energy (12),
where the crystal anisotropy ($v$-) term is also taken into
account. Here we shall present our basic results for the effect of
the Cooper pair anisotropy on the FS phase; the crystal anisotropy
is neglected ($v=0$).

The dimensionless anisotropy parameter $w=\bar u/(u + \bar u)$ can
be either positive or negative depending on the sign of $\bar u$.
Obviously when $\bar u > 0$, the parameter $w$ will be positive
too ($0<$ w$<1$). We shall illustrate the influence of Cooper-pair
anisotropy in this case. The order parameters ($M$, $\phi_j$,
$\theta_j$) are given by Eqs.~(40), (43),
 \begin{equation}
\label{eq54}
\phi^2=\frac{\pm \gamma M-r-\gamma_1 M^2}{(1-w)} \ge 0\:,
\end{equation}
and
\begin{equation}
\label{eq55}
(1- w - \gamma_1^2)M^3 \pm \frac{3}{2} \gamma \gamma_1 M^2
+\left[t(1-w)-\frac{\gamma^2}{2}-\gamma_1 r\right]M \pm \frac{\gamma
r}{2}=0\:,
\end{equation}
where the meaning of the upper and lower sign is the same as
explained just below Eq.~(43). We consider the FS domain
corresponding to the upper sign in the Eq.~(54) and (55). The
stability conditions for FS read,
\begin{equation}
\label{eq56}
\frac{ (2-w)\gamma M- r -\gamma_1M^2}{1-w} \ge 0\:,
\end{equation}
\begin{equation}
\label{eq57}
\frac{1-2w}{1-w}(\gamma M -wr-w \gamma_1 M^2) > 0\:,
\end{equation}
and
\begin{equation}
\label{eq58}
\frac{1}{1-w}\left[3(1-w-\gamma_1^2) M^2 + 3 \gamma \gamma_1 M +
t(1-w)-\frac{\gamma^2}{2} -\gamma_1 r \right]\geq 0\:.
\end{equation}
For $M\ne (\gamma/2 \gamma_1)$ we can express the function $r(M)$
defined by Eq.~(54), substitute the obtained expression for $r(M)$
in the existence and stability conditions (54)-(57) and do the
analysis in the same way as for $w=0$. The calculations show that
in the domain $r>0$, FS is stable for $w<0.5$, when $w=0.5$ there
is a marginal stability, and for $w>0.5$ the FS-phase is unstable
($0<w<1$).

The results can be used to outline the phase diagram and calculate the
thermodynamic quantities. This is performed in the way explained in
the preceding Sections.

{\bf 7. Conclusion}

We have done an investigation of the M-trigger effect in
unconventional ferromagnetic superconductors. This effect due to
the $M\psi_1\psi_2$-coupling term in the GL free energy consists
of bringing into existence of superconductivity in a domain of the
phase diagram of the system that is entirely in the region of
existence of the ferromagnetic phase. This form of coexistence of
unconventional superconductivity and ferromagnetic order is
possible for temperatures above and below the critical temperature
$T_s$, which corresponds to the standard phase transition of
second order from normal to Meissner phase -- usual uniform
superconductivity in a zero external magnetic field, which appears
outside the domain of existence of ferromagnetic order. Our
investigation has been mainly intended to clarify the
thermodynamic behaviour at temperatures $T_s< T < T_f$, where the
superconductivity cannot appear without the mechanism of
M-triggering. We have described the possible ordered phases (FM
and FS) in this most interesting temperature interval.

The Cooper pair and crystal anisotropies have also been
investigated and their main effects on the thermodynamics of the
triggered phase of coexistence have been established. In
discussions of concrete real material one should take into account
the respective crystal symmetry but the variation of the essential
thermodynamic properties with the change of the type of this
symmetry is not substantial when the low symmetry and low order
(in both  $M$ and $\psi$) $\gamma$-term is present in the free
energy.

Below the superconducting critical temperature $T_s$ a variety of
pure superconducting and mixed phases of coexistence of
superconductivity and ferromagnetism exists and the thermodynamic
behavior at these relatively low temperatures is more complex than
in known cases of improper ferroelectrics. The case $T_f < T_s$
also needs a special investigation.

Our results are referred to the possible uniform superconducting
and ferromagnetic states. Vortex and other nonuniform phases need
a separate study.

The relation of the present investigation to properties of real
ferromagnetic compounds, such as UGe$_2$, URhGe, and ZrZn$_2$, has
been discussed throughout the text. In these real compounds the
ferromagnetic critical temperature is much larger than the
superconducting critical temperature $(T_f \gg T_s)$ and that is
why the M-triggering of the spin-triplet superconductivity is very
strong. Moreover, the $\gamma_1$-term is important to stabilize
the FM order up to the absolute zero (0 K), as is in the known
spin-triplet ferromagnetic superconductors. The
neglecting~\cite{Walker:2002} of the symmetry conserving
$\gamma_1$-term prevents the description of the known real
substances of this type. More experimental information about the
values of the material parameters ($a_s, a_f, ...$) included in
the free energy (12) is required in order to outline the
thermodynamic behavior and the phase diagram in terms of
thermodynamic parameters $T$ and $P$. In particular, a reliable
knowledge about the dependence of the parameters $a_s$ and $a_f$
on the pressure $P$, the value of the characteristic temperature
$T_s$ and the ratio $a_s/a_f$ at zero temperature are of primary
interest.

{\bf Acknowledgments:}

 DIU thanks the hospitality of
MPI-PKS-Dresden. Financial support by SCENET (Parma) and JINR (Dubna) is also
acknowledged.

\end{document}